\documentclass[a4paper,twocolumn,11pt,accepted=2026-01-31]{quantumarticle}
\pdfoutput=1
\usepackage[utf8]{inputenc}
\usepackage[english]{babel}
\usepackage[T1]{fontenc}
\usepackage[numbers]{natbib}

\usepackage[colorlinks=true, linkcolor=blue, citecolor=blue, filecolor=blue, urlcolor=blue]{hyperref}

\usepackage{enumerate}
\usepackage{amsfonts,amssymb,amsmath}
\usepackage[]{graphics,graphicx,epsfig}
\usepackage{amsthm}
\usepackage{mathbbol}
\usepackage{graphicx}
\usepackage{dcolumn}
\usepackage{natbib}
\usepackage{color}
\usepackage{multirow}
\usepackage{ulem}
\usepackage{bm}
\usepackage{url}
\usepackage{float}
\usepackage{enumitem}
\usepackage{lipsum}
\usepackage[justification=justified,format=plain]{caption}
\usepackage{subcaption}
\usepackage[dvipsnames]{xcolor}
\usepackage{xurl}

\allowdisplaybreaks[1]

\usepackage{tikz}
\usepackage{environ}
\usetikzlibrary{automata,arrows,positioning,calc}
\usetikzlibrary{quantikz}

\newcommand{\Past}        {  {\overleftarrow{X}} }
\newcommand{\past}        {  \overleftarrow{x} }
\newcommand{\Future}      {  \overrightarrow{X}}
\newcommand{\future}      {  \overrightarrow{x}}

\newcommand{\PastFuture}  {  \overleftrightarrow{X} }

\def\ket#1{\left|#1\right\rangle}
\def\bra#1{\left\langle#1\right|}
\def\braket#1{\left\langle#1\right\rangle}
\newcommand{\norm}[1]{\left\lVert#1\right\rVert}
\newcommand{\tr}[1]{\text{Tr}\left[#1\right]}

\begin{document}

\title{Ideal stochastic process modeling with post-quantum quasiprobabilistic theories}

\author{Kelvin Onggadinata}
\email{kelvin.onggadinata@ntu.edu.sg}
\thanks{Current address: School of Physical and Mathematical Sciences, Nanyang Technological University, Singapore.}
\affiliation{Centre for Quantum Technologies,
National University of Singapore, 3 Science Drive 2, Singapore 117543, Singapore}
\affiliation{Department of Physics,
National University of Singapore, 3 Science Drive 2, Singapore 117543, Singapore}

\author{Andrew Tanggara}
\email{andrew.tanggara@gmail.com}
\affiliation{Centre for Quantum Technologies,
National University of Singapore, 3 Science Drive 2, Singapore 117543, Singapore}
\affiliation{Nanyang Quantum Hub, School of Physical and Mathematical Sciences, Nanyang Technological University, Singapore 637371, Singapore}

\author{Mile Gu}
\email{mgu@quantumcomplexity.org}
\affiliation{Nanyang Quantum Hub, School of Physical and Mathematical Sciences, Nanyang Technological University, Singapore 637371, Singapore.}
\affiliation{Centre for Quantum Technologies,
National University of Singapore, 3 Science Drive 2, Singapore 117543, Singapore}
\affiliation{Centre for Quantum Technologies,
Nanyang Technological University, Singapore 637371, Singapore}
\affiliation{Majulab, CNRS-UNS-NUS-NTU International Joint Research Unit, UMI No. 3654, Singapore 117543, Singapore}

\author{Dagomir Kaszlikowski}
\email{phykd@nus.edu.sg}
\affiliation{Centre for Quantum Technologies,
National University of Singapore, 3 Science Drive 2, Singapore 117543, Singapore}
\affiliation{Department of Physics,
National University of Singapore, 3 Science Drive 2, Singapore 117543, Singapore}

\begin{abstract} 
In stochastic modeling, the excess entropy -- the mutual information shared between a process's past and future -- represents the fundamental lower bound of the memory needed to simulate its dynamics. However, this bound cannot be saturated by either classical machines or their enhanced quantum counterparts. Simulating a process fundamentally requires us to store more information in the present than is shared between the past and the future. Here, we consider a generalization of hidden Markov models beyond classical and quantum models, referred to as n-machines, that allow for negative quasiprobabilities. We show that under the collision entropy measure of information, the minimal memory of such models can equal the excess entropy. Our results suggest that negativity can be a useful resource for achieving nonclassical memory advantage. 
\end{abstract}

\maketitle

\section{Introduction}\label{sec: introduction}

Modeling of stochastic time-series is pervasive in many fields of quantitative science. The most commonly used framework is the {\it hidden Markov model} (HMM) \cite{rabiner1986introduction,vidyasagar2014hidden} which has been demonstrated to be useful in many ranges of application, such as, speech recognition \cite{jelinek1998statistical}, dynamical spin systems \cite{crutchfield1997statistical,suen2017classical}, machine learning \cite{ghahramani1995factorial,fine1998hierarchical}, neuroscience \cite{haslinger2010computational} and stock markets \cite{yang2008increasing}. Needless to say, understanding the properties and capabilities of HMM is of essential interest.

In many situations (e.g., language modeling, stock markets), these models are designed for prediction. Such models are causal, that is, their internal states contain no oracular information --- information about the future beyond knowing the entire past~\cite{crutchfield2010synchronization}. Such models require a memory amount of $C$, which is bounded below by the {\it excess entropy} $\mathbf{E}$ --- the mutual information shared between a process's past and its future; the rationale being that any violation would violate the data processing inequality. The ideal model, as motivated by Occam's Razor, would be one where $C = \mathbf{E}$, such that all information a model tracks is reflected in the future. It turns out, however, this is generally impossible. The provably minimal classical causal models --- the  $\epsilon$-machine, exhibit memory costs $C_\mu > \mathbf{E}$ for many processes; motivating $C_\mu$ as a distinct measure of process complexity that captures how hard it is to simulate a process in a given causal direction~\cite{crutchfield1989inferring,shalizi2001computational}. 

Could one find a more memory-efficient model by dropping the causality requirement, considering the most general machines that can store oracular information? Finding the memory-minimal machines in this more general setting remains an open problem, but specific case studies  \cite{lohr2009generative,lohr2009non,lohr2012predictive,ruebeck2018prediction} have demonstrated the capacity to use less memory than $C_\mu$. Nevertheless, such machines still require memory strictly greater than $\mathbf{E}$~\cite{thompson2018causal}. 

A promising avenue is to consider using a quantum system to encode the model. First developed in \cite{gu2012quantum} and subsequently studied in \cite{mahoney2016occam,riechers2016minimized,binder2018practical,liu2019optimal,elliott2020extreme,elliott2022quantum}, they are known as q-machines and belong to a subclass of general hidden quantum Markov model (HQMM) \cite{wiesner2008computation,monras2012hidden,monras2016quantum}. Such q-machines provably use memory $C_q$ strictly less than $C_\mu$, whenever $C_\mu > \mathbf{E}$. Yet such quantum machines remain non-ideal. Such non-ideal models dissipate heat during operation, and thus the physical universe appears to demand fundamental irreversibility when it comes to stochastic simulation~\cite{loomis2020thermal}. This motivates the main question of this work: Can we find an ideal model, and what are the constraints we can drop to achieve this?

\begin{figure}[!htb]
    \centering
    \includegraphics[width=0.95\linewidth]{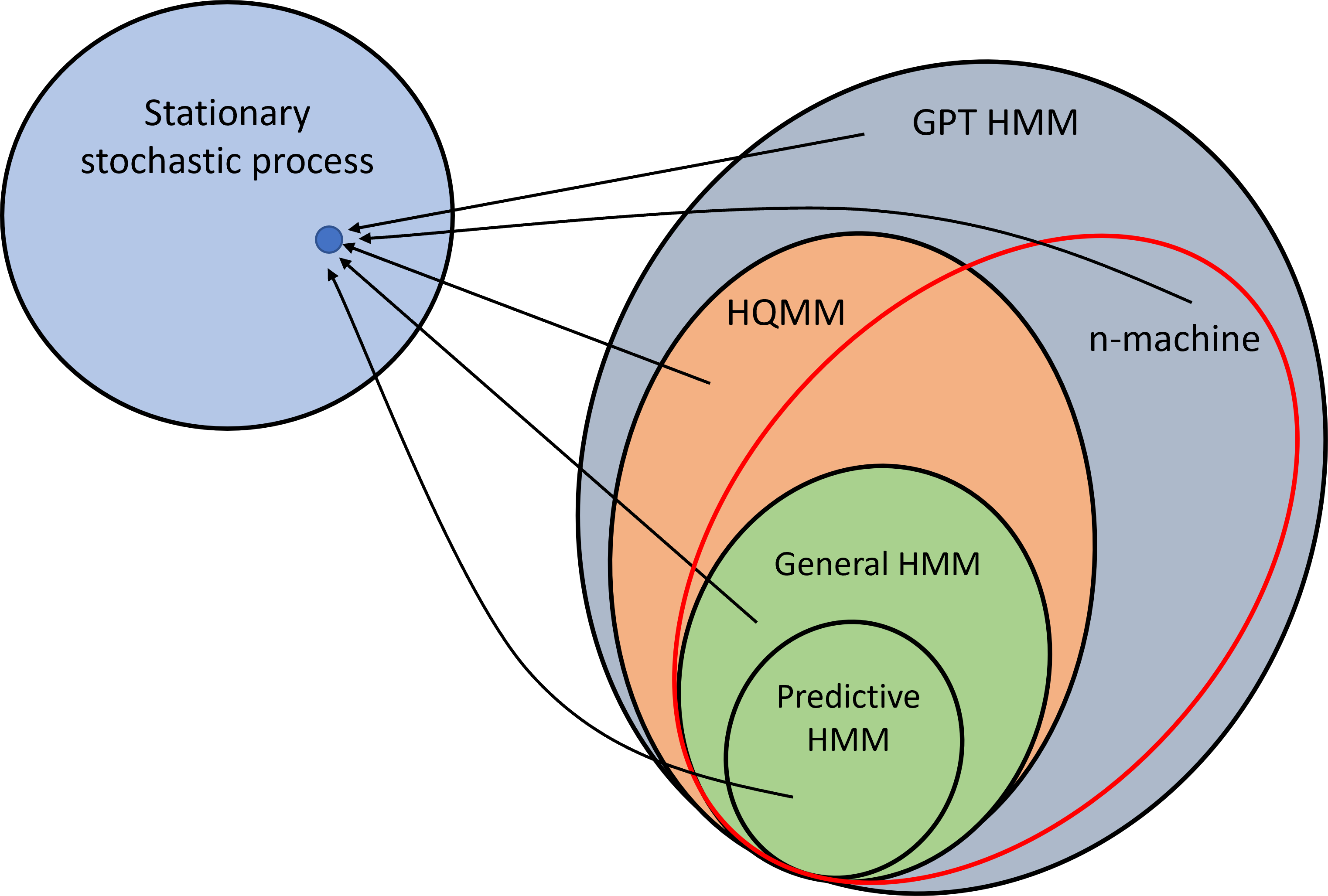}
    \caption{Hierarchy of different models generating the same stochastic process. As discussed in the main text, the sets of n-machine (circled in red) belong to a subset of GPT HMM.}
    \label{fig: hmm hierarchy}
\end{figure}

In this article, we explore relaxation of positivity in the context of stochastic process modeling. We provide a general protocol to construct such an ideal model by allowing negative probability (or quasiprobability) when memory is measured by R{\'e}nyi entropy of order $\alpha=2$ (also known as collision entropy). Quasiprobability is an unavoidable feature in any quantum theories \cite{ferrie2010necessity}, and more importantly, has been attributed as a resource for ``quantum advantage'' in the field of quantum information (nonlocality \cite{abramsky2014operational,onggadinata2023simulations}, contextuality \cite{spekkens2008negativity,booth2022contextuality}) and quantum computation \cite{veitch2012negative,howard2014contextuality,pashayan2015estimating,kaszlikowski2021little}. Furthermore, it encompasses theories beyond quantum, such as generalized probability theory (GPT) \cite{barrett2008information}, and thus provides insights into information-theoretic properties that can be advantageous. With this in mind, the proposed ideal model here can be thought of as a ``GPT formulation of HMM". A similar notion has also been studied in \cite{vidyasagar2014hidden} and more recently in \cite{fanizza2024quantum, dong2023promotion}. In \cite{fanizza2024quantum}, they showed a stochastic process generated by GPT HMM but not realizable by any hidden quantum Markov model (HQMM) in finite dimensions. The difference here is that we consider only stochastic processes with finite-dimensional classical and quantum realizations, and show how to construct the GPT HMM representation and compare its memory usage with that of other classical or quantum models (see Fig. \ref{fig: hmm hierarchy}). Ref. \cite{dong2023promotion} looked into the application of Markov model with negative transition probability for the modeling of gene interaction.

This article is sectioned as follows. In Section \ref{sec: preliminaries}, we review the classical, quantum, and general theory of HMM representations of stochastic processes, and the information-theoretic measures used commonly in the literature. Then, in Section \ref{sec: alternative measures}, we will introduce the alternative information-theoretic measures that we use and show that they are well-defined and operationally meaningful, even when extended to quasiprobabilities. Section \ref{sec: n machine} contains the main result of this paper, where we describe the protocol to construct the ideal generative machine and discuss its properties. We then illustrate this with a few examples in Section \ref{sec: examples} and study the relationship between negativity and memory advantage. We conclude with a general discussion and suggestions for future work.

\section{Preliminaries}\label{sec: preliminaries}

\subsection{Stochastic process}\label{sec: stochastic process}

We consider a discrete-valued, discrete-time stochastic process that is defined as a sequence of random variables $X_t$, where at each time $t$ the system emits a realization $x_t$ from some discrete alphabet $\mathcal{A}$. At each time $t$, the sequence can be partitioned into two infinite strings, namely the past $\Past_t \equiv \dots X_{t-2}X_{t-1}$ and the future $\Future_t\equiv X_tX_{t+1}\dots$. The pattern of a stochastic process could then be defined as a joint probability distribution over the bi-infinite strings $P(\Past_t,\Future_t)$. For past and future of finite length $L$ we denote them as $\Past_t^L \equiv X_{t-L}\dots X_{t-1}$ and $\Future^L \equiv X_{t}\dots X_{t+L-1}$, respectively. The outputs with length $L$ is referred as word $w \equiv \future_t^L = x_tx_{t+1}\dots x_{t+L-1} \in \mathcal{A}^L$, where $\mathcal{A}^L$ is the language containing all length-$L$ realizations. To simplify our case, we only consider stationary processes such that the statistics are independent of time $t$ and henceforth, we will drop the subscript $t$. A key property of any stochastic process is its {\it excess entropy}
\begin{eqnarray}
\mathbf{E} & \equiv & I[\Past;\Future] \nonumber \\
&=& \sum_{\past,\future}P(\past,\future)\log\frac{P(\past,\future)}{P(\past)P(\future)} \label{eq: excess entropy}
\end{eqnarray}
defined as the mutual information\footnote{In this article, all logarithm is computed to base 2, such that $\log 2 = 1$ bit.} shared between $\Past$ and $\Future$. The excess entropy, sometimes also referred to as predictive information, is the  information about the future that is contained in past observations.

A hidden Markov model (HMM) is a finite state machine with a set of internal states $\bm{\mathcal{S}}$, and transition probabilities $T_{j;k}^{(x)}$ that describes the probability a machine in state $S_j \in \bm{\mathcal{S}}$ at time $t$ will emit output $x$ and transition to state $S_k$. After repeated operation, each model will generate some sequence of outputs $\Past,\Future$ governed by some stochastic process $P(\Past,\Future)$. The task of modeling then corresponds to identifying a suitable HMM (i.e., $\bm{\mathcal{S}}$ and $T_{j;k}^{(x)}$) that generates $P(\Past, \Future)$. In general, each stochastic process has many possible models that can be grouped into different classes.

\subsection{Classical predictive and generative models}\label{sec: hmm}

Predictive models are a prominent sub-class of hidden Markov machine. They assume there exists some deterministic map that takes each possible past $\past$ to a corresponding state $s_{\past} \in \mathcal{S}$, such that $P(\Future|\mathcal{S} = s_{\past})=P(\Future|\Past =\past)$. Satisfying this condition then gives a systematic method for us to encode each past into the state of the predictive model machine, such that the conditional future behaviour is statistically identical to that of the machine. Any such machine then displays the property of causal shielding --- the internal states of the machine can render the past and future to be independent such that $P(\Past,\Future|\mathcal{S}) = P(\Future|\mathcal{S})P(\Past|\mathcal{S})$. 

The field of computational mechanics deals with finding the simplest predictive models. In particular, they have singled out $\epsilon$-machine as the memory-minimal predictive model. The basic idea of $\epsilon$-machine is to look for sets of pasts with the same future statistics \cite{shalizi2001computational}. Such pasts that share an equivalence relation
\begin{equation}\label{eq: causal equivalence relation}
    \past \sim_{\epsilon} \past' \Longleftrightarrow P(\Future|\Past=\past)=P(\Future|\Past=\past')
\end{equation}
are grouped into a causal state $\sigma_k = \epsilon(\past)$. We then denote the set of all causal states as $\bm{\mathcal{S}}=\{\sigma_1,\dots\}$ and $\mathcal{S}$ as the causal state's random variable. 

Governing how the states evolve with time and the probabilities of generating the next instance's realization is the {\it symbol-labeled state transition probabilities}:
\begin{equation}\label{eq: symbol labelled transition matrices}
    T^{(x)}_{j;k} \equiv P(\mathcal{S} = \sigma_k, \Future^{L=1} = x|\mathcal{S} = \sigma_j)\, ,
\end{equation}
where $\sigma_j$ and $\sigma_k$ refer to the initial and final causal state, respectively, and $x$ is the symbol emitted after the transition takes place. The set of all transition matrices is denoted as $\bm{\mathcal{T}} = \{T^{(x)}\,\vert \, T^{(x)}_{j;k}= P(\sigma_k,x|\sigma_j)\}_{x\in \mathcal{A}}$ with each $T^{(x)}$ being a substochastic map. Due to the deterministic partitioning by $\sim_\epsilon$, $\epsilon$-machines have an important property known as {\it unifilarity} (sometimes referred to as deterministic). That is, for symbol $x$ emitted from state $\sigma_j$, there is at most only one end state $\sigma_k$ of such transition to occur. The unifilarity plays a key role as it underlies direct calculation to several important properties of a process \cite{shalizi2001computational}. The stationary distribution for the causal states,
\begin{equation}
\pi \equiv  [\pi_1, \dots,\pi_{|\bm{\mathcal{S}}|}] = [P(\sigma_1),\dots,P(\sigma_{|\bm{\mathcal{S}}|})]\, ,
\end{equation}
can be obtained by solving the eigenproblem
\begin{equation}\label{eq: stationary distribution eigenproblem}
    \pi T = \pi\, ,
\end{equation}
where $T = \sum_{x\in\mathcal{A}}T^{(x)}$ is known as the state transition matrix. Note that the transition matrix is row-stochastic, $\sum_{k=1}^{|\bm{\mathcal{S}}|} T_{j;k} = 1$ for all $j$. Thus, it also follows that the stationary distribution is also stochastic, i.e., $\sum_{k=1}^{|\bm{\mathcal{S}}|} \pi_k = 1$. To summarize, an $\epsilon$-machine is characterized by the tuple $(\mathcal{A}, \bm{\mathcal{S}}, \bm{\mathcal{T}},\pi)$. Generating a word $w=x_0\dots x_{L-1}$ of a stochastic process can then be easily described using components of the $\epsilon$-machine following
\begin{equation}\label{eq: classical word realization}
P(w) = \pi T^{(w)}\bm{1} = \pi T^{(x_0)}T^{(x_1)}\dots T^{(x_{L-1})}\bm{1}
\end{equation}
where $\bm{1} = [1,1,\dots, 1]^T$.

An $\epsilon$-machine can quantify the amount of past information required to store in order to faithfully predict future outcomes. Namely, due to the equivalence relation in Eq. \eqref{eq: causal equivalence relation}, knowing the state of the $\epsilon$-machine is as good as knowing the entire past. We can quantify this amount of memory using the R\'enyi-$\alpha$ entropy (or $\alpha$-entropy) on the stationary distribution
\begin{equation}
C_\mu^{(\alpha)} \equiv H_\alpha[\pi] = \frac{1}{1-\alpha} \log\sum_k \pi_k ^\alpha\, , \label{eq: renyi statistical complexity}
\end{equation}
where $\alpha \in [0,\infty)$, which we refer to as its $\alpha$-statistical complexity \cite{crutchfield1989inferring,shalizi2001computational}.  Out of all possible $\alpha$, there are two that are particularly interesting and widely used:
\begin{itemize}
    \item ($\alpha=0$) {\it Topological complexity} \cite{crutchfield1994observing}:
    \begin{equation}
        C_\mu^{(0)} = \dim |\bm{\mathcal{S}}|\, . \label{eq: topological complexity}
    \end{equation}
    This has the interpretation of single-shot memory cost, and is quantified simply as the number of causal states.
    \item ($\alpha\to 1$) {\it Statistical complexity} \cite{crutchfield1989inferring}:
    \begin{equation}
        C_\mu \equiv C_\mu^{(1)} = -\sum_k \pi_k \log \pi_k\, . \label{eq: statistical complexity}
    \end{equation}
    This quantifies the average amount of past information stored in the causal states using Shannon entropy. Alternatively, it has also been interpreted as the amount of information required to communicate via a classical channel to synchronize two predictive models \cite{mahoney2016occam}.
\end{itemize}
From Jensen's inequality, we have the relation $C_\mu^{(0)} \geq C_\mu$.

$\epsilon$-machines, like all other predictive models, are a subset of \textit{causal models} --- models whose internal states $\mathcal{S}$ do not contain information about the future that is not already available from the past. This information is quantified by the oracular information $I[\mathcal{S};\Future|\Past]$ and is zero for $\epsilon$-machines \cite{crutchfield2010synchronization}. In this sense, knowing the states of $\epsilon$-machine is as good as knowing the entire past but no better, which also implies that $\mathbf{E} = I[\mathcal{S};\Future]$. Meanwhile, $\epsilon$-machines are also memory-minimal among all causal models. As such $C_\mu$ and $C_\mu^{(0)}$ are considered intrinsic properties of a stochastic process --- quantifying fundamentally how difficult a stochastic process is to predict or model in a given temporal direction. They have been used to understand temporal structure in diverse domains, from neuroscience and stock markets to self-organisation and identifying the transition to chaos \cite{crutchfield1997statistical, suen2017classical, ghahramani1995factorial, fine1998hierarchical, haslinger2010computational, yang2008increasing}. In general we have $C_\mu \geq \mathbf{E}$ with the equality achieved iff the $\epsilon$-machine is reversible (for each end state $\sigma'$ and symbol $x$, there is at most only one state $\sigma$ making the transition) but this is rarely the case \cite{gu2012quantum}.

Beyond causal models, there exist more general HMMs that could possess lower memory than the $\epsilon$-machines \cite{lohr2009generative,lohr2009non,lohr2012predictive,ruebeck2018prediction}. These models are referred as generative-machines (g-machines) and are characterized as potentially having non-zero oracular information. An example is the $\epsilon$-machine of a time-reversed process when viewed in reverse time \cite{ellison2011information}. The quantification of the memory for g-machine takes the same form as $\epsilon$-machine. To differentiate them we shall use the subscript $g$, and so Eqs. \eqref{eq: topological complexity}-\eqref{eq: statistical complexity} becomes $C_\mu^{(0)} \to C_g^{(0)}$ and $C_\mu \to C_g$, where they are understood to be the minimal memory corresponding to an optimal g-machine. $C_g$ also has been referred to as generative complexity \cite{ruebeck2018prediction}. 

Since generative machines encompass predictive machines, we have the general relation\cite{crutchfield2009time}:
\begin{equation}
    C_\mu \geq C_g \geq \mathbf{E}
\end{equation}
However, there is no systematic means to identify generative complexity; as such, $C_g$ is known only for selective case studies. In these case studies, $C_g$ remains strictly greater than $\mathbf{E}$.

\subsection{Hidden quantum Markov model}\label{sec: hqmm}

\begin{figure}[!htb]
    \centering
    \includegraphics[width=0.9\columnwidth]{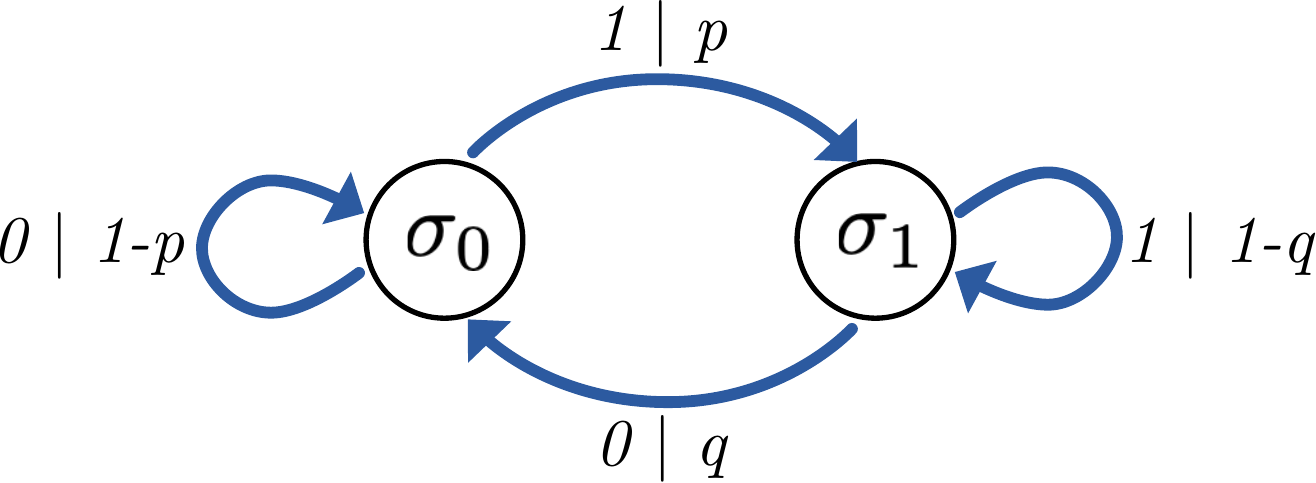}
    
    \vspace{1.5em}
    
    \includegraphics[width=1\columnwidth]{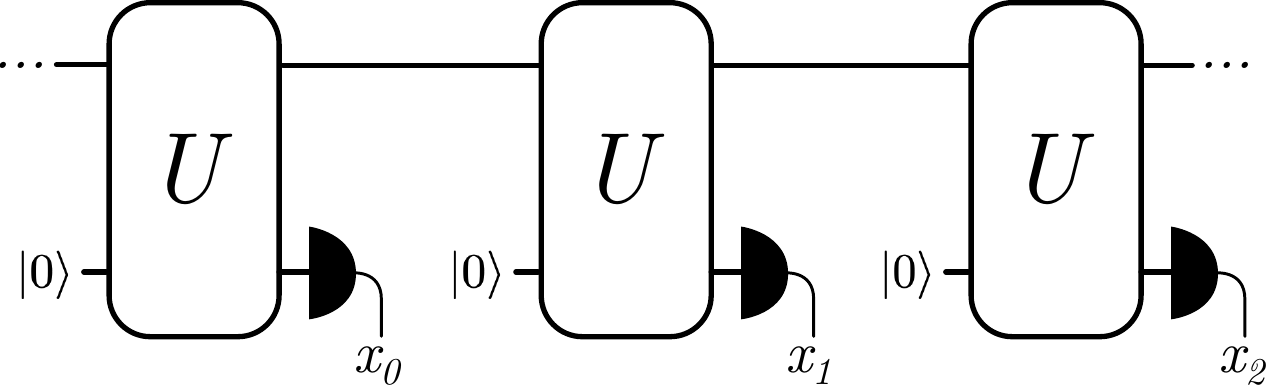}
    \caption{A hidden quantum Markov model (bottom figure) can be constructed from a classical HMM (top figure).
    In this illustration, two states $\sigma_0,\sigma_1$ of the HMM are encoded into quantum states $|\sigma_0\rangle,|\sigma_1\rangle$ (which may be non-orthogonal) in which the action of the unitary $U$ of the hidden quantum Markov model (HQMM) is defined (see Eq.~\eqref{eq: q machine unitary}).
    The hidden memory state of the HQMM is illustrated by the top wire, which propagates the unitaries over time.
    At time step $t$, an ancilla $|0\rangle$ is fed into unitary $U$ along with the memory state, then a subsystem at the output of $U$ is measured to output the realization $x_t$.
    }
    \label{fig: quantum_HMM}
\end{figure}

Traditionally, $\epsilon$-machines are considered the optimal causal models. However, with many instances where going quantum has shown an advantage, Ref. \cite{gu2012quantum} showed that their memory can be further reduced when the machine state is encoded within a quantum system. Here, we describe the protocol to construct a quantum $\epsilon$-machine (q-machine) following the construction in \cite{mahoney2016occam,riechers2016minimized,binder2018practical} and illustrate it in Fig. \ref{fig: quantum_HMM}.

Starting from the $\epsilon$-machine $(\mathcal{A},\bm{\mathcal{S}}, \bm{\mathcal{T}},\pi)$, the q-machine constructs a set of internal states $\bm{\mathcal{S}}_q = \{\ket{\sigma_k}\}_k$, where each element can be seen as the quantization of the classical causal states. To make predictions, the q-machine applies a unitary to the internal state coupled with an ancilla state, say $\ket{0}$, and emits the output upon the unitary action. Repeated application of resetting the ancilla state and applying the same unitary generates a sequence of output systems --- whose observation of the computational basis exhibits identical statistics to the original stochastic process. Reference \cite{binder2018practical} showed that such a unitary can always be found, and they obey
\begin{equation}\label{eq: q machine unitary}
 U\ket{\sigma_j}\ket{0} = \sum_{x\in\mathcal{A}}\sum_{k}\sqrt{T^{(x)}_{j;k}} \ket{\sigma_k}\ket{x}\, .
\end{equation}
The stationary state of the q-machine is then an ensemble of those internal states:
\begin{equation}\label{eq: q machine state}
    \rho = \sum_k \pi_k\ket{\sigma_k}\bra{\sigma_k}\, .
\end{equation}
In general, the overlap $\braket{\sigma_j|\sigma_k}$ ($j\neq k)$ are nonzero, which has been attributed as the source for quantum memory advantage.

Analogous to the classical $\alpha$-statistical complexity, the memory retained by $\rho$ can be quantified by the quantum R\'enyi-$\alpha$ entropy:
\begin{equation}\label{eq: quantum renyi statistical complexity}
    C_q^{(\alpha)} \equiv S_{\alpha}[\rho] = \frac{1}{1-\alpha}\log\tr{\rho^\alpha}\, .
\end{equation}
Thus, we have the quantum analogy of Eqs. \eqref{eq: topological complexity}-\eqref{eq: statistical complexity}:
\begin{eqnarray}
C_q^{(0)} &=& \log\left(\text{rank}\left[\rho\right]\right)\, , \label{eq: quantum topological complexity} \\
C_q \equiv C_q^{(1)} &=& S_{\text{vN}}[\rho] \, ,\label{eq: quantum statistical complexity}
\end{eqnarray}
where $S_{\text{vN}}[\bullet]$ is the von Neumann entropy \cite{nielsen2010quantum}. Following this construction, it has been shown in general that $C_q \leq C_\mu$ with equality saturated iff the classical $\epsilon$-machine is already ideal ($C = \mathbf{E}$) \cite{gu2012quantum}. The reduction in the memory requirement by the quantum simulator represents the quantum advantage over the classical model. However, $C_q > \mathbf{E}$ is found for most processes \cite{mahoney2016occam}, and in certain instances, such quantum constructions are provably optimal~\cite{thompson2018causal}. This implies that the best quantum predictor is still not ideal as its memory requirement is still greater than $\mathbf{E}$.

Up to this point, we have described the quantum enhancement of $\epsilon$-machines. For g-machine, a quantum enhancement takes a similar form and is discussed in more depth in \cite{elliott2021memory}. Similarly to the classical case, we also write $C_{qg}^{(0)}$ and $C_{qg}^{(1)} \equiv C_{qg}$ for the topological complexity and generative complexity of the quantum g-machine, respectively. Although the construction in \cite{elliott2021memory} showed that $C_{qg} \leq C_{g}$, the specific machines constructed did not possess memory smaller than $C_{q}$.

\subsection{Generalized probability theory of hidden Markov model}\label{sec: quasi-realization}

Generalized theories of hidden Markov model attempt to contextualize both classical and quantum hidden Markov models into a larger set of theories for models capable of stochastic process realization. If the realization of a stochastic process by quantum theory is achieved via a quantum state evolving under a completely-positive dynamics, then one can consider a (quasi-)realization of the same stochastic process through a state vector evolving under a linear map. To make the definition more precise, we provide a brief description of generalized theories of HMM (also referred as quasi-realization theories) following \cite{vidyasagar2014hidden, fanizza2024quantum}. 

We define a quasi-realization of a stationary stochastic process to be a quadruple $(\mathcal{V}, \pi, D, \tau)$. Here, $\mathcal{V}$ is a vector space with dual space $\mathcal{V}^*$, $\tau \in \mathcal{V}$, and $\pi \in \mathcal{V}^*$.
Furthermore, we define a linear map $D: \mathcal{A}^* \to \mathcal{L}(\mathcal{V})$ to be a unital representation on the set of all finite words $\mathcal{A}^*\equiv \bigcup_{l\geq 0}\mathcal{A}^l$ over $\mathcal{V}$ if the condition
\begin{equation}\label{eq: quasi-realization transition matrices}
    D^{(\emptyset)} = \mathbb{1}\, ,\quad D^{(\vec{x})}D^{(\vec{y})} = D^{(\vec{x}\vec{y})}\,, \quad \forall\, \vec{x},\vec{y}\in \mathcal{A}^*\, ,
\end{equation}
holds.
Here $\emptyset$ is the empty word belonging to $\mathcal{A}^{L}$ when $L=0$. Note that $\mathcal{L}(\mathcal{V})$ is the set of linear operators acting on $\mathcal{V}$ and we say that $D$ is unital if it preserves the identity element. Moreover, we have the following relation
\begin{equation}\label{eq: quasi-realization left right eigenvector}
\pi\left[\sum_{x\in \mathcal{A}}D^{(x)}\right] = \pi\, ,\quad \left[\sum_{x \in \mathcal{A}}D^{(x)}\right]\tau = \tau\, ,
\end{equation}
and the generation of the stochastic process for some word $w$ is obtained through
\begin{equation}
P(w) = \pi D^{(w)}\tau\, ,\quad \forall\, w\in\mathcal{A}^*\, .
\end{equation}
The above relation represents the generalized form of Eq. \eqref{eq: classical word realization}.

From here, one can notice that both the classical HMM and HQMM are subtheories of the quasi-realization theory. In particular, classical HMM corresponds to the \textit{positive realization} with the quadruple $(\mathbb{R}^d,\pi, T, \mathbf{1})$, where $\mathbb{R}^d$ is a $d$-dimensional real vector space, $T^{(x)}$ are non-negative transition matrices defined in Eq. \eqref{eq: symbol labelled transition matrices}, $\pi\in\left(\mathbb{R}^d\right)^*$ is the stationary distribution and $\tau = \mathbf{1} \equiv [1,1,\dots, 1]^T$ as seen in Eq. \eqref{eq: stationary distribution eigenproblem} and \eqref{eq: classical word realization}. These are exactly the components of the HMM that we have seen in Sec. \ref{sec: hmm}. 
Meanwhile, HQMM corresponds to the \textit{completely positive realization} with the quadruple $(\mathcal{B}(\mathcal{H})^{\text{sa}}, \tr{\rho(\cdot)}, \mathcal{E}, \mathbb{1})$, where $\mathcal{B}(\mathcal{H})$ is the space of bounded operators on some finite-dimensional Hilbert space $\mathcal{H}$, $\mathcal{B}(\mathcal{H})^{\text{sa}}$ is the vector space of all self-adjoint operators, $\mathcal{E}^{(x)}$ are completely-positive maps, $\tr{\rho(\cdot)}\in (\mathcal{B}(\mathcal{H})^{\text{sa}})^*$ is the stationary state of the unital map $\sum_{x}\mathcal{E}^{(x)}$, and $\tau = \mathbb{1}$ is identity operator on $\mathcal{B}(\mathcal{H})$. 
Again, we can see that the components in Sec. \ref{sec: hqmm} fit into this subtheory, where $\sum_{x}\mathcal{E}^{(x)}$ is obtained from tracing out the ancilla subsystem from the unitary in Eq. \eqref{eq: q machine unitary} and $\rho$ is given in Eq. \eqref{eq: q machine state}. 
The probability of a word $w = x_0\dots x_{L-1}$ is then given by $\tr{\rho \mathcal{E}^{(w)}{}^\dag(\mathbb{1})}$.
Later, we shall see that the construction of our ideal generative model is an instance of quasi-realization. 

\section{Alternative information-theoretic measures}\label{sec: alternative measures}

In the previous section, we saw several definitions of information-theoretic measures, such as statistical complexity and excess entropy, that are primarily defined using Shannon or von Neumann entropy in the quantum domain, where $\alpha \to 1$. This definition works very well and is preferred, as it has been studied extensively and has many desirable properties, making it convenient to calculate and relate to one another. However, to prepare for the upcoming construction, we will need an alternative definition to those information-theoretic measures. This is done because we anticipate the appearance of quasiprobability in objects such as stationary distributions and transition matrices. In this case, Shannon entropy is no longer compatible with quasiprobability distributions. As such, in this section, we will introduce an alternative definition of several information-theoretic measures seen above, ensuring they remain well-behaved for quasiprobabilities. Moreover, we will show that this definition works well even for previous models such as $\epsilon$-machines and q-machines, capturing most of the important properties and relationships between the measures. 

\subsection{R{\'e}nyi statistical complexity}

Starting with the measure of the memory required by the machine's internal states, recall that they can be quantified generally with $\alpha$-statistical complexity as defined in Eq. \eqref{eq: renyi statistical complexity}. In this work, we shall pick $\alpha=2$ as our choice of measure. Referring it simply as 2-statistical complexity, it reads
\begin{equation}\label{eq: 2 statistical complexity}
    C_\mu^{(2)} \equiv H_2[\pi] = -\log\sum_{k=1}^{|\bm{\mathcal{S}}|} \pi_k^2\, .
\end{equation}
In some literature, the R{\'e}nyi-2 entropy is also colloquially known as \textit{collision entropy} \cite{bosyk2012collision}. Similarly for the quantum case, we also pick $\alpha=2$ for Eq. \eqref{eq: quantum renyi statistical complexity} and denote it as quantum 2-statistical complexity:
\begin{equation}
    C_q^{(2)}(\rho) \equiv S_2[\rho] = -\log \tr{\rho^{2}}\, .
\end{equation}
Compared to the previous measure, it has the relation $C_\mu \geq C_\mu^{(2)}$ and $C_q \geq C_q^{(2)}$. 

\begin{figure}[!htb]
    \centering
    \includegraphics[width=1\linewidth]{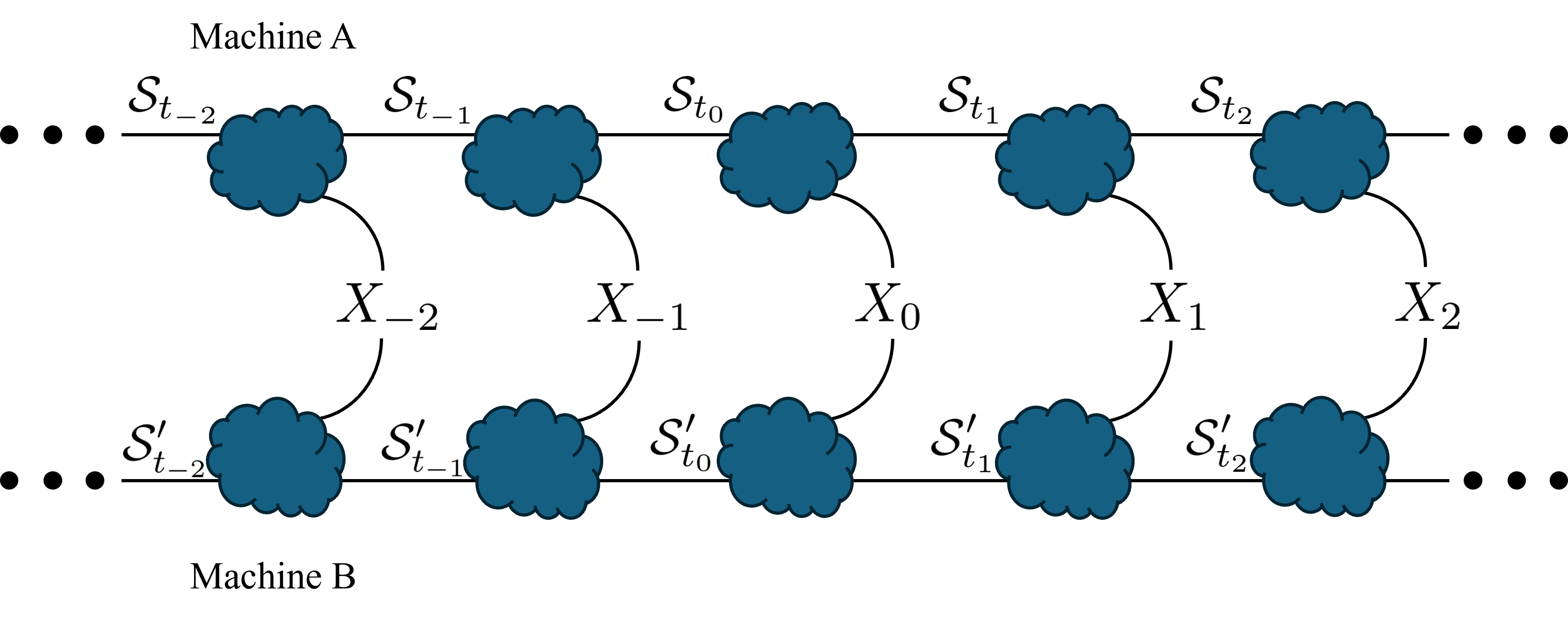}
    \caption{Illustration of the simulation/generation task as synchronization of two identical and independent machines' internal states.}
    \label{fig: simulation task}
\end{figure}

In \cite{crutchfield1989inferring}, $\alpha$-statistical complexity has been proposed as the measure of information processing capacity of an $\epsilon$-machine. However, as mentioned before, much of the focus is placed on $\alpha=0$ and $\alpha=1$. Here, we provide some operational meaning to 2-statistical complexity in the setting of stochastic process modeling. We assume that there are 2 identical and independent machines, referred as machine A and B, that generate the same stochastic process $P(\Past,\Future)$ (see Fig. \ref{fig: simulation task}). If we want machine A and machine B to have identical output statistics at an arbitrary time, then we require that the two machines' internal states match at that time. As such, for any given time $t$, the amount of uncertainty that the internal state of machine A and machine B matches is given by
\begin{equation}
-\log P(\mathcal{S}=\mathcal{S}') = -\log \sum_{k}\pi_k^2\, ,
\end{equation}
which is exactly the 2-statistical complexity as defined in Eq. \eqref{eq: 2 statistical complexity}. Here, we denote $\mathcal{S}$ and $\mathcal{S}'$ as the random variables for internal states of machine A and B, respectively, and we have dropped the time subscript due to the stationarity assumption. The quantity above also represents the amount of information overhead we need to store to simulate or generate the stochastic process correctly at any arbitrary time. When the process is modeled using quantum theory, then it is natural to use the quantum R{\'e}nyi-2 entropy as the machine's internal states are described using density operators.

Moreover, the 2-statistical complexities can also be understood operationally in the context of the source-coding/compression task of $N$ i.i.d. sequence of symbols from an arbitrary binary random source $X$ with probability $P_X(x)$ into $NR$ bits.
Namely, a sequence of $N$ independent random variables $X_1\dots X_N$ with $X_i$ distributed identically as $X$ is encoded to a random variable $Z$ over $NR$ symbols $\{1,\dots,2^{NR}\}$ by a conditional probability $P_N^E(Z|x_1\dots x_N)$ corresponding to the encoder.
Thus, the probability of string $x_1\dots x_N$ sampled from $X_1\dots X_N$ successfully recovered by a decoder, modeled by a conditional probability $P_N^D(X_1\dots X_N | z)$, is therefore
\begin{equation}
\begin{aligned}
    P_\mathrm{suc}(x_1\dots x_N) &= \sum_z P_N^D(x_1\dots x_N | z) \\
    &\quad \times P_n^E(z|x_1\dots x_N) \;.
\end{aligned}
\end{equation}
So the decoding success probability is given by 
\begin{equation}
\begin{aligned}
    P_\mathrm{suc} &= \sum_{x_1\dots x_N,z} P_N^D(x_1\dots x_N | z) \\
    &\quad \times P_N^E(z|x_1\dots x_N) P_X(x_1)\dots P_X(x_N) \;.
\end{aligned}
\end{equation}
Here, $H_2[X]$ acts as a cut-off on the encoding rate $R$ indicating a bound for an encoding to not suffer an arbitrarily large decoding error as $N\rightarrow\infty$ \cite[Theorem 1]{csiszar1995generalized}.
More precisely, decoding error probability $P_\mathrm{err} = 1-P_\mathrm{suc}$ is lower-bounded as
\begin{align}
    P_\mathrm{err} > 1-2^{-\frac{1}{2}N (H_2[X]-R) + o(N)} \;.
\end{align}
Namely, the decoding error rate is always non-zero and approaches one as $N\rightarrow\infty$ with rate  $2^{-\frac{1}{2}N (H_2[X]-R) + o(N)}$ if we compress each instance of $X$ into less than $H_2[X]$ bits.
In modeling a stochastic process with some model $\mathcal{R}$, its complexity $H_2[\mathcal{R}]$ can simply be thought of as a lower-bound on the necessary amount of bits one can compress $\mathcal{R}$ in order to not have arbitrarily large decoding error, i.e. as the machine attempts to emulate future statistics.

Another operationally meaningful interpretation of $H_2$ is from the cryptographic task of privacy amplification, which is an information transmission from Alice to Bob using a given $n$-bits source of randomness $X$ and an $r$-bit encoding $g:\{0,1\}^n\rightarrow\{0,1\}^r$ in the presence of an eavesdropper Eve.
It was shown in \cite{bennett1995generalized,renner2008security} that $H_2[X]$ determines the safety of this transmission from Eve.
More precisely, for a protocol that randomly chooses $g$ that is known to Eve, the amount of security of encoded transmission $g(X)$ from Eve as quantified by $H_2$ is
\begin{align}
    H_2[G(X)|G] \geq r-\log\Big(1+2^{r-H_2[X]}\Big)\, ,
\end{align}
where $G$ is the random variable for the encoding function $g$.
Hence, one may interpret $H_2[X]$ as the information content of $X$ by noting that $H_2[X]$ bounds the amount of information that can be extracted from $X$ by some observer (in this case, Eve) who only possesses knowledge of the extraction protocol $G$, but no knowledge of $X$.

\subsection{R{\'e}nyi excess entropy}

As for the excess entropy Eq. \eqref{eq: excess entropy} that quantifies the amount of information shared between the past and future, we need a generalized notion of R\'enyi-mutual information. To this extent, we use the $\alpha$-mutual information as introduced in \cite{csiszar1972class,sibson1969information,csiszar1995generalized} defined as 
\begin{equation}\label{eq: hv alpha mutual information}
    I_{\alpha}[X;Y] \equiv \frac{\alpha}{\alpha-1}\log\sum_{y}\left[\sum_{x}P(x)\left(P(y|x)\right)^{\alpha}\right]^{\frac{1}{\alpha}}
\end{equation}
for any Markov chain $X\to Y$ and any $\alpha\in[0,\infty)$. At $\alpha \to 1$, it recovers the Shannon mutual information, and the properties are studied extensively in \cite{ho2015convexity,verdu2015alpha}. From there, a relevant result for us is Theorem 5 of \cite{ho2015convexity}, which shows that
\begin{equation}
    H_\alpha[X] \geq I_{\frac{1}{\alpha}}[X;Y]\, .
\end{equation}
Due to the choice of $\alpha=2$ for the R{\'e}nyi statistical complexity, it is then natural to define
\begin{eqnarray}
\mathbf{E}_{\frac{1}{2}} &\equiv &I_{\frac{1}{2}}[\Past;\Future] \nonumber \\
&=& -\log\sum_{\future\in\Future} \left(\sum_{\past\in\Past}P(\past)\sqrt{P(\future|\past)}\right)^2 \label{eq: 1/2 excess entropy}
\end{eqnarray}
as $\frac{1}{2}$-excess entropy quantifying the information known about the future by the past. 

The R\'enyi-$\frac{1}{2}$ mutual information has a special operational interpretation as a channel-coding ``cut-off" rate of a discrete memoryless channel (DMC) decoding error probability $P_\mathrm{err}$ \cite{gallager1965simple,arimoto1973converse,csiszar1995generalized,polyanskiy2010arimoto}. More precisely for a DMC $P_{Y|X}$ (forming a Markov chain $X\rightarrow Y$) and some input code-word probability distribution $P_X$, the decoding error probability $P_\mathrm{err}$ is bounded by
\begin{align}\label{eq:renyi_channel_coding_error_bound_rate}
    P_\mathrm{err} \leq 2^{-N(I_\frac{1}{2}[X;Y] - R)}
\end{align}
for a length $N$ block code with rate $R$ (i.e. an encoding of bit string message $\{0,1\}^{NR}$ to $N$ i.i.d. transmissions of code-word $X$). Hence for a stochastic process $\PastFuture$, if we consider $P(\Future|\Past)$ as a channel and $P(\Past)$ as a distribution over code-words, then the excess entropy $\mathbf{E}_\frac{1}{2} = I_\frac{1}{2}[\Past;\Future]$ is simply the least number of bits that one can encode into $\Past$ that one can reliably recover from $\Future$ with vanishing error with an increasing number of copies of $P(\Future|\Past)$.
Similarly for a predictive model $\mathcal{S}$, the $\frac{1}{2}$-excess entropy $\mathbf{E}_\frac{1}{2} = I_\frac{1}{2}[\mathcal{S};\Future]$ signifies the least number of bits that can be encoded into $\mathcal{S}$ and then reliably recovered from $\Future$.

\subsection{Relationship between R\'enyi statistical complexity and R\'enyi excess entropy}

In the following, we show that several important properties of the $\epsilon$-machine or q-machine still apply under these new measures. Firstly, analogous to $I[\mathcal{S};\Future]=\mathbf{E}$, we show that $I_{\frac{1}{2}}[\mathcal{S};\Future]=\mathbf{E}_{\frac{1}{2}}$ with the proof delineated in Appendix \ref{sec: appendix proof new measures}. Therefore, under this definition, the $\epsilon$-machine remains predictive and the data processing inequality is still satisfied. One can also easily infer that this is also true for q-machine.

Secondly, we also found that $C_\mu^{(2)} \geq C_q^{(2)} \geq \mathbf{E}_{\frac{1}{2}}$ analogous to the important relation $C_\mu \geq C_q \geq \mathbf{E}$. The proof can be found in Appendix \ref{sec: appendix proof new measures}. This generalizes the notion of excess entropy as a lower bound for the entropic memory measure for the classical and quantum models. For processes with non-ideal classical or quantum models, we conjecture that there exists a model in GPT with memory saturating this lower bound. Interestingly, the qualitative behaviors of the statistical complexities and the excess entropy remain largely intact under the new definitions. This will be much more apparent in the examples shown in Sec. \ref{sec: examples}. In addition to that, we also look into Even Process \cite{crutchfield2003regularities} where observation that $C_\mu = C_q = \mathbf{E}$ is still found to be the same here, i.e., $C_\mu^{(2)} = C_q^{(2)} = \mathbf{E}_{\frac{1}{2}}$. The Even Process is not studied further in the subsequent section, since no futher memory advantage can be made.

Lastly, an important result by Ref. \cite{crutchfield2009time} showed that $\mathbf{E} = I[\Past;\Future] = I[\mathcal{S}^+; \mathcal{S}^-]$, where $\mathcal{S}^+ = \mathcal{S} = \epsilon^+(\Past)$ is the causal states for the {\it forward-direction} stochastic process and $\mathcal{S}^- = \epsilon^-(\Future)$ is the causal states for the {\it reverse-direction} stochastic process \cite{ellison2009prediction}. This provides a convenient way to calculate $\mathbf{E}$, especially for highly cryptic processes. The generalization of this to our measure here is however not true in general, i.e., $\mathbf{E}_{\frac{1}{2}}$ is not equal to $I_{\frac{1}{2}}[\mathcal{S}^+; \mathcal{S}^-]$ for all stochastic processes. This is simply due to the non-symmetric nature of $\alpha$-mutual information used here, that is, $I_{\alpha}[X;Y] \neq I_{\alpha}[Y;X]$. However, we will show that for some processes, they can be equal. In Appendix \ref{sec: appendix proof new measures}, we show their equivalence for the two stochastic processes we studied in Sec. \ref{sec: examples}.

\subsection{Collision entropy as an information measure for quasiprobabilities}

As mentioned before, the main purpose of using R\'enyi entropy is to accommodate quasiprobabilities. In this case, we have focused on the use of $H_2$ as a quantifier for memory. This function is well-defined on the set of quasiprobabilities with nonzero components $\mathcal{Q} = \{q \in \mathbb{R}^N\, |\, \sum_{k=1}^{N}q_k = 1,\, q_k\neq 0\}\}$, for which it satisfies 
\begin{enumerate}
    \item {\it Real-valuedness}: $H_2[q] \in \mathbb{R}$.
    \item {\it Symmetric} on the components of $q$.
    \item {\it Normalization}: $H_2[(\frac{1}{N}, \dots , \frac{1}{N})] = \log N$.
    \item {\it Continuous} on any quasiprobability $q \in \mathcal{Q}$.
    \item {\it Continuously differentiable}: The derivative of $H_2[q]$ exists for each $q_k$ and continuous for all $q\in \mathcal{Q}$.
    \item {\it Additive} on the products of quasiprobabilities: $H_2[q \otimes q'] = H_2[q] + H_2[q']$.
    \item {\it Mean-value property}: For any $q,q'\in \mathcal{Q}$, there exists a strictly monotonic and continuous function $g$ such that 
    \begin{equation}
        H[q\cup q'] = g^{-1}\left[g(H[q']) + g(H[q])\right]\, ,
    \end{equation}
    where $H[q\cup q'] = H[(q_1,\dots, q_N, q'_1, \dots q'_N)]$.
    \item {\it Schur-concave}: For $q,q' \in \mathcal{Q}$ such that $q \succ q'$ ($q$ majorizes $q'$), then $H_2[q] \leq H_2[q']$.
\end{enumerate}
The properties 2-7 are taken directly from the original list of axioms proposed by R\'enyi \cite{renyi1961measures} (later proven to be true by Daroczy \cite{daroczy1963gemeinsame}), which fully characterize the family of R\'enyi entropy. Property 1 is also taken from the aforementioned reference but is modified from {\it nonnegativity} to {\it real-valuedness} to accommodate quasiprobabilities and the requirement of entropy to be a real-valued function. These properties are proven to be satisfied in Ref. \cite{brandenburger2025axiomatization} for $\alpha=2k$, where $k$ takes positive integer. The eighth property then further reinforces that $H_2$ can be used as a measure of disorder between quasiprobability distributions. Ref. \cite{koukoulekidis2022constraints} proved this by first considering a generalized notion of relative majorization for quasiprobabilities, which can be consistently shown by `masking' the negativity in the quasiprobability with some reference distribution. This follows with establishing the Schur-concavity of R\'enyi entropy with $\alpha = \frac{2a}{2b-1}$, where $a,b$ are integers satisfying $a\geq b$. With all of these in mind, we pick $\alpha=2$ --- belonging to both the family of $\alpha$'s in \cite{brandenburger2025axiomatization} and \cite{koukoulekidis2022constraints} --- to put in use here. It is also noteworthy to point out that $\alpha=2$ has been the natural and preferred choice in the subject of axiomatization of quantum theory involving information-theoretic principles \cite{wootters2007discrete,brukner2009information, brandenburger2022renyi, onggadinata2023qubits}.

Moreover, we argue that $H_2[q]$ is a reasonable measure of information contained in the quasiprobability distribution $q$ from a more operational perspective by reverting to classical simulation of quasiprobability sampling \cite{abramsky2014operational,pashayan2015estimating} which has found applications in the study of classical simulations of quantum computation \cite{pashayan2015estimating,rahimi2016sufficient,koukoulekidis2022constraints,koukoulekidis2022faster}, quantum error-correction \cite{temme2017error,takagi2021optimal,takagi2022fundamental}, classical simulation of quantum memory channels \cite{yuan2021universal}, and local simulation of non-local quantum channels \cite{mitarai2021overhead}.
This quasiprobability sampling technique can be described generally by a classical probability $p = (p_i)_i$, where $p_i = \frac{|q_i|}{\sum_j |q_j|}$.
A simple observation can be made by taking the R\'enyi-2 entropy of the classical probability distribution $p$ and decomposing it as
\begin{align}\label{eq: collision entropy simulation split terms}
\begin{split}
    H_2[p] &= -\log\sum_i p_i^2 \\
    &= \underbrace{-\log\Big(\sum_i |q_i|^2\Big)}_{\textup{(i)}} + \underbrace{2\log\Big(\sum_j |q_j|\Big)}_{\textup{(ii)}} \;,
\end{split}
\end{align}
where term (i) is precisely $H_2[q]$, and term (ii) is a logarithmic factor of the amount of negativity present in $q$.
The latter quantity is precisely what has been identified as a resource for quantum computational advantage known as {\it mana} \cite{veitch2014resource}, whereas the quantity inside the log (the sum negativity) is also a common measure of nonclassicality and has been shown as the overhead of classical simulation of quasiprobability sampling \cite{pashayan2015estimating,koukoulekidis2022constraints}.
This simple observation shows that the entropy of a quasiprobability $q$ is simply part of how much information is contained in its classical simulation $p$ subtracted by the amount of overhead cost of running this simulation, hence giving us the information content of $q$.

\section{Ideal generative model}\label{sec: n machine}

\subsection{Construction protocol}

In this section, we describe the main result of the paper. That is, we provide a general method to construct the ideal generative model, which we shall refer as negative-machine (n-machine). Given a stochastic process $P(\Past,\Future)$, the protocol to construct an n-machine is as follows.

{\it Inputs:} $\epsilon$-machine $(\mathcal{A},\bm{\mathcal{S}}, \bm{\mathcal{T}}, \pi)$ representation of stochastic process $P(\Past,\Future)$.

{\it Outputs:} n-machine with internal states $\tilde{\bm{\mathcal{S}}}$ and transition probabilities $\tilde{\bm{\mathcal{T}}}$.  

\begin{enumerate}
    \item Construct the internal states for the n-machine through the following extension: $\sigma_k \to \{\tilde{\sigma}_{k,l_k}\}_{l_k}$ where $l_k\in L_k$. The set $L_k$ need not have the same size for different $k$. The new set of states has now become
    \begin{equation}
        \tilde{\bm{\mathcal{S}}} = \bigcup_k \, \{\tilde{\sigma}_{k,l_k}\}_{l_k\in L_k}\, .
    \end{equation}
    \item After extending the set of internal states, we need to establish the transition probabilities between the states $\tilde{\bm{\mathcal{T}}} = \{\tilde{T}^{(x)} \,\vert \, \tilde{T}^{(x)}_{j,l_j;k,l_k} = P(\tilde{\sigma}_{k,l_k},x|\tilde{\sigma}_{j,l_j})\}_{x\in\mathcal{A}}$. We constrain the transition probabilities to adhere to
    \begin{equation}\label{eq: n-machine construct 1}
        \sum_{l_k}P(\tilde{\sigma}_{k,l_k},x|\tilde{\sigma}_{j,l_j}) = P(\sigma_{k},x|\sigma_{j})\quad \forall \, k,j,l_j,x\, .
    \end{equation}
    \item Minimize $C_n^{(2)}\equiv H_2[\tilde{\mathcal{S}}]$ such that $C_n^{(2)} \geq \mathbf{E}_{\frac{1}{2}}$.
\end{enumerate}

\begin{figure}[!htb]
    \centering
    \includegraphics[width=1\linewidth]{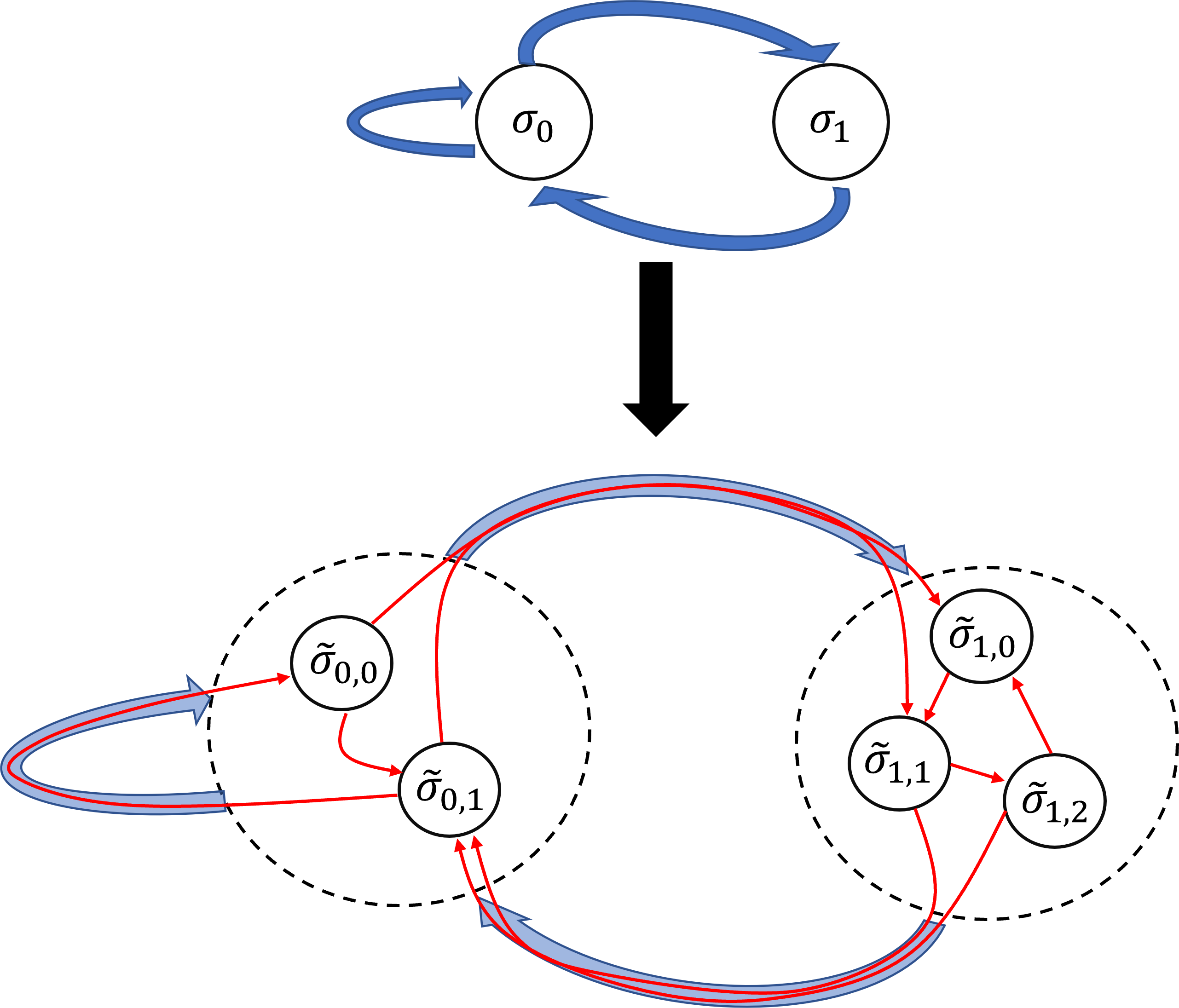}
    \caption{Simple illustration of the protocol. Starting from the $\epsilon$-machine (top diagram), the n-machine (bottom diagram) is constructed by copying the causal states. The transition arrows in red are the extended transitions allowing quasiprobabilities with their net effects still emulating the previous transitions (in blue) of the $\epsilon$-machine.}
    \label{fig: n machine cartoon}
\end{figure}

The main idea behind the protocol above is that the behavior or purpose of the causal states $\sigma_k$ is captured completely by each $\tilde{\sigma}_{k,l_k}$. By preserving the properties of each causal state and the total probability of outputting symbols, this protocol constructs an HMM that still generates an identical stochastic process. Meanwhile, the additional degrees of freedom through the extension of the states allows one to reduce the state memory if one considers negative transition probabilities. The encoding to n-machine and its mechanisms are illustrated in Fig. \ref{fig: n machine cartoon}.

As an example, suppose that we have $\bm{\mathcal{S}}=\{\sigma_0,\sigma_1\}$. From Step 1, we then could have $\sigma_0 \to \{\tilde{\sigma}_{0,0}, \tilde{\sigma}_{0,1}\}$ and $\sigma_1\to \{\tilde{\sigma}_{1,0}, \tilde{\sigma}_{1,1},\tilde{\sigma}_{1,2}\}$, where here $l_0\in L_0=\{0,1\}$ and $l_1\in L_1=\{0,1,2\}$. The extended set of states become $\tilde{\bm{\mathcal{S}}}=\{\tilde{\sigma}_{0,0}, \tilde{\sigma}_{0,1}\}\cup\{\tilde{\sigma}_{1,0}, \tilde{\sigma}_{1,1},\tilde{\sigma}_{1,2}\}=\{\tilde{\sigma}_{0,0}, \tilde{\sigma}_{0,1},\tilde{\sigma}_{1,0}, \tilde{\sigma}_{1,1},\tilde{\sigma}_{1,2}\}$. Then, from Step 2, we have new free parameters such as $P(\tilde{\sigma}_{0,0},x|\tilde{\sigma}_{0,0})$, $P(\tilde{\sigma}_{0,1},x|\tilde{\sigma}_{0,0})$, $P(\tilde{\sigma}_{0,0},x|\tilde{\sigma}_{0,1})$, $P(\tilde{\sigma}_{0,1},x|\tilde{\sigma}_{0,1})$ that can be chosen to obey
\begin{eqnarray}
    P(\sigma_0, x| \sigma_0) &=& P(\tilde{\sigma}_{0,0},x|\tilde{\sigma}_{0,0}) + P(\tilde{\sigma}_{0,1},x|\tilde{\sigma}_{0,0}) \nonumber \\
    &=& P(\tilde{\sigma}_{0,0},x|\tilde{\sigma}_{0,1}) + P(\tilde{\sigma}_{0,1},x|\tilde{\sigma}_{0,1})\, .\nonumber
\end{eqnarray}
With the transition probabilities now allowed to take negative values, the transition matrix $\tilde{T} = \sum_{x\in\mathcal{A}}\tilde{T}^{(x)}$ has been generalized into a quasi-stochastic matrix (matrix with real values but sums to 1 in each row). Lastly, Step 3 seeks to minimize the memory of the n-machine as quantified by 2-statistical complexity:
\begin{equation}
C_n^{(2)} \equiv H_2[\tilde{\mathcal{S}}] = -\log \sum_{k,l_k}\tilde{\pi}_{k,l_k}^2\, ,
\end{equation}
where
\begin{eqnarray}
\tilde{\pi} &\equiv & [\tilde{\pi}_{0,0}, \tilde{\pi}_{0,1}, \dots ,\tilde{\pi}_{1,0},\dots] \nonumber \\ 
&=& [P(\tilde{\sigma}_{0,0}),P(\tilde{\sigma}_{0,1}),\dots, P(\tilde{\sigma}_{1,0}),\dots] \nonumber
\end{eqnarray}
is the quasi-stochastic stationary distribution for the internal states satisfying Eq. \eqref{eq: quasi-realization left right eigenvector}. Note that $C_n^{(2)}$ can be smaller than $\mathbf{E}_{\frac{1}{2}}$ when there is too much negativity in the model. Hence, the minimization of $C_n^{(2)}$ is lower bounded by $\mathbf{E}_{\frac{1}{2}}$ as it acts as the limitation we imposed on any generative model. We draw a comparison of this limitation as analogous to the bound imposed in Bell nonlocality experiments for any nonclassical model \cite{brunner2014bell}.\footnote{For instance, in the CHSH experiment \cite{clauser1969propesed}, classical and quantum theory could reach the value of 2 and $2\sqrt{2}$, respectively, but there exist nonlocal systems, such as Popescu-Rohrlich box \cite{popescu1994quantum}, that could reach the maximal value of 4. The maximal value is imposed by the no-signalling principle and the algebraic limit of the correlation functions. One could easily find more nonlocal systems that go beyond, but they are cut off by the physical limit.}

Here, the stationary distribution for the n-machine's states is found in the same manner as it is obtained for the classical case in Eq. \eqref{eq: stationary distribution eigenproblem}. Since $\tilde{T}$ is a quasi-stochastic matrix, one can always find a left eigenvector with eigenvalue 1, which then corresponds to the stationary distribution. Note, however, that the stationary distribution has now been extended to a quasiprobability distribution. This might be surprising but we point out that the q-machine's state $\rho$ can also be represented as a state quasiprobability distribution under the quasiprobability representation (QPR) of quantum mechanics \cite{ferrie2008frame,ferrie2009framed,ferrie2011quasi}. We believe that quasiprobability also plays a role in the memory reduction in quantum model, and this warrants a separate and dedicated study elsewhere.

\subsection{Properties of n-machine}

We now discuss some of the consequences and properties of the n-machine. Firstly, following the construction above will relax the unifilarity condition of $\epsilon$-machine, thus n-machine is non-unifilar in general. In fact, the construction can also be done starting from any (non-unifilar) generative model. However, due to the lack of a constructive method for generative model, it is more reliable to start from the $\epsilon$-machine representation.

Secondly, the new stationary distribution $\tilde{\pi}$ is reducible to its previous stationary distribution $\pi$. More specifically, it has a neat property that the probability of the extended states $\{\tilde{\sigma}_{k,l_k}\}_{l_k}$ is the same as the original state $\sigma_k$:
\begin{equation}\label{eq: n-machine construct 3}
    \sum_{l_k}P(\tilde{\sigma}_{k,l_k}) = P(\sigma_k) \quad \forall \, k\, .
\end{equation}
Proof is in Appendix \ref{sec: appendix proof n-machine}. This observation is akin to the fact that coarse-graining measurements on non-classical states often lead to classical behavior in the macroscopic limit \cite{kofler2007classical}.

Thirdly, it is easy to see from Eq. \eqref{eq: n-machine construct 1} that we have
\begin{equation}\label{eq: n-machine construct 2}
    P(x|\tilde{\sigma}_{k,l_k}) = P(x|\sigma_{k})\quad \forall\, x,k,l_k\,.
\end{equation}
This tells us that at any given time, the probability of seeing an observable $x$ from the copied states $\{\tilde{\sigma}_{k,l_k}\}_{l_k}$ is the same as the original state $\sigma_k$. Moreover, since $P(x|\sigma_k)$ is nonnegative, this also ensures that the probability of generating observable $x$ is always nonnegative, and so no negative probability is observed in the outcome. As a matter of fact, this can be extended to 
\begin{equation}\label{eq: n-machine construct 2 extended}
    P(\future|\tilde{\sigma}_{k,l_k}) = P(\future|\sigma_k) \quad \forall\, x,k,l_k\, .
\end{equation}
From here, it is easy to see that the generation of any stochastic process by the n-machine is exactly the same as its $\epsilon$-machine, i.e, they generate an identical stochastic process. Moreover, this is also indicative by the fact that n-machine's states still fully capture the past-future information, i.e., $I_{\frac{1}{2}}[\tilde{\mathcal{S}};\Future] = \mathbf{E}_{\frac{1}{2}}$. Again, the proofs are found in Appendix \ref{sec: appendix proof n-machine}.

The n-machine constructed here can easily be seen to belong to the quasi-realization as described in Sec. \ref{sec: quasi-realization}. In particular, its quadruple is given by $(\mathbb{R}^d,\tilde{\pi}, \tilde{\mathcal{T}}, \mathbf{1})$, where we have considered $\mathbb{R}^d$, the vector space of finite dimension $d$ over real numbers, $\tilde{\mathcal{T}}^{(x)}$ are quasi-stochastic matrices, and $\tilde{\pi}$ is the quasi-stochastic stationary distribution. It is important to note that the stochastic process studied here can be thought of as being independent of any theory. The model in which we can simulate/generate them, however, can be drawn from particular theories that entails different memory complexities (see Fig. \ref{fig: hmm hierarchy}).

In the next section, we demonstrate our n-machine construction for several examples of stochastic process, and show that it is possible to achieve the maximal memory compression with it.

\section{Examples}\label{sec: examples}

\subsection{Perturbed Coin Process}\label{sec: example perturbed coin}

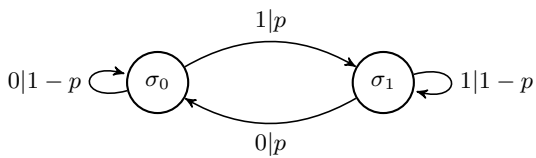
\begin{figure}[!htb]
    \centering
    \begin{tikzpicture}[->, >=stealth', auto, semithick, node distance=3cm]
    \tikzstyle{every state}=[fill=white,draw=black,thick,text=black,scale=1]
    \node[state]    (A)                     {$\sigma_0$};
    \node[state]    (B)[right of=A]   		{$\sigma_1$};
    \path
    (A) edge[loop left] 		node{$0|1-p$}    (A)
        edge[bend left, above] 	node{$1|p$}      (B)
    (B) edge[loop right]  		node{$1|1-p$}    (B)
        edge[bend left, below] 	node{$0|p$} 	 (A);
    \end{tikzpicture}
    \caption{$\epsilon$-machine representation of Perturbed Coin Process. The nodes represent the internal states of the model and the edges labelled $x|p$ represent the transition made between states with probability $p$ while emitting symbol $x$. The same notation is used throughout in subsequent diagrams.}
    \label{fig: perturbed coin epsilon machine}
\end{figure}

The first example we will look at is the Perturbed Coin Process ($0<p<1, p\neq \frac{1}{2})$ with the $\epsilon$-machine representation as shown in Fig. \ref{fig: perturbed coin epsilon machine}. The $\frac{1}{2}$-excess entropy can be easily calculated to be $\mathbf{E}_\frac{1}{2} = 1 - 2\log\left(\sqrt{p} + \sqrt{1-p}\right)$. The stationary distribution is given by $\pi = [1/2, 1/2]$ so the 2-statistical complexity is $C_\mu^{(2)} = 1$. Note that for $p=\frac{1}{2}$, the process reduces to an Unbiased Coin Process where it is an independent, identically distributed (IID) process with a single causal state $\epsilon$-machine representation. Hence at $p=\frac{1}{2}$, $C_\mu^{(2)}$ discontinuously reduces to zero.

The q-machine representation has been studied in Ref. \cite{gu2012quantum}. In particular, the quantum states are expressed as $\ket{\sigma_0} = \sqrt{1-p}\ket{0} + \sqrt{p}\ket{1}$ and $\ket{\sigma_1} = \sqrt{p}\ket{0} + \sqrt{1-p}\ket{1}$. With the encoding to the quantum state as $\rho = \frac{1}{2}(\ket{\sigma_0}\bra{\sigma_0} + \ket{\sigma_1}\bra{\sigma_1}$, the quantum 2-statistical complexity reads $C_q^{(2)} = -\log\left(\frac{1}{2} + 2p(1-p)\right)$.

\begin{figure}[!htb]
\centering
    \begin{subfigure}[t]{0.99\linewidth}
    \centering
    \begin{tikzpicture}[->, >=stealth', auto, semithick, node distance=3cm]
    \tikzstyle{every state}=[fill=white,draw=black,thick,text=black,scale=1]
    \node[state]    (A)                     {$\sigma_A$};
    \node[state]    (B)[right of=A]   		{$\sigma_B$};
    \path
    (A) edge[loop left] 		node{$1|\frac{1-2p}{1-p}$}    (A)
        edge[bend left, above] 	node{$1|\frac{p}{1-p}$}      (B)
    (B) edge[loop above]  		node{$1|\frac{p^2}{1-p}$}    (B)
        edge[loop below]  		node{$0|1-p$}    (B)
        edge[bend left, below] 	node{$1|\frac{p(1-2p)}{1-p}$} 	(A);
    \end{tikzpicture}
    \caption{}
    \label{fig: perturbed coin rjmc model left}
    \end{subfigure}
    \begin{subfigure}[t]{0.99\linewidth}
    \centering
    \begin{tikzpicture}[->, >=stealth', auto, semithick, node distance=3cm]
    \tikzstyle{every state}=[fill=white,draw=black,thick,text=black,scale=1]
    \node[state]    (A)                     {$\sigma_A$};
    \node[state]    (B)[right of=A]   		{$\sigma_B$};
    \path
    (A) edge[loop above] 		node{$1|1-p$}    (A)
        edge[loop below] 		node{$0|1-p$}    (A)
        edge[bend left, above] 	node{$1|2p-1$}   (B)
    (B) edge[bend left, below] 	node{$0|1$} 	 (A);
    \end{tikzpicture}
    \caption{}
    \label{fig: perturbed coin rjmc model right}
    \end{subfigure}
    \caption{RJMC's g-machine representation of Perturbed Coin Process for (a) $0< p < \frac{1}{2}$ and (b) $\frac{1}{2} < p < 1$.}
    \label{fig: perturbed coin rjmc model}
\end{figure}
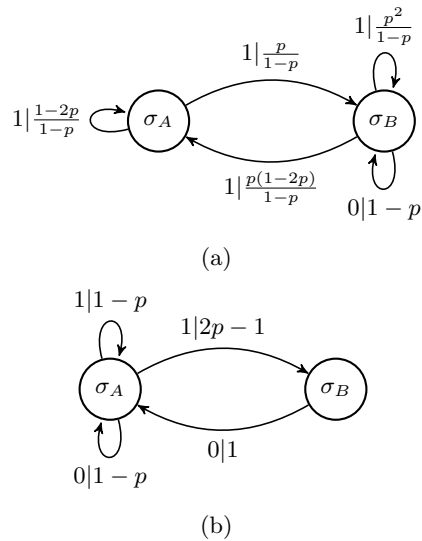

\begin{figure}[!htb]
\centering
\includegraphics[width=\linewidth]{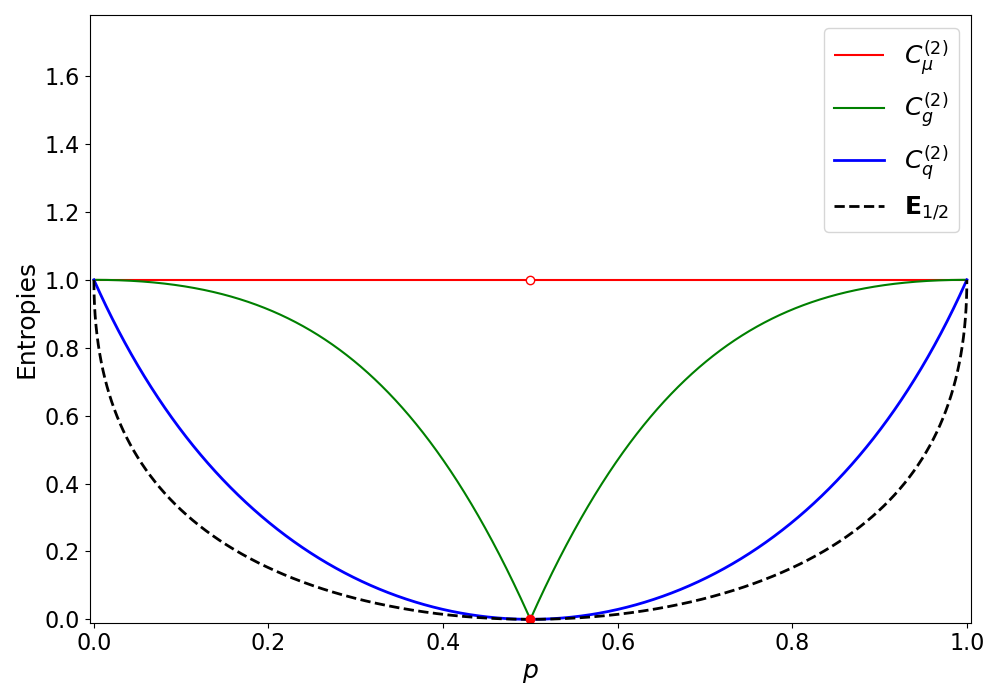}
\caption{Comparison of the memory usage by the different models for Pertubed Coin Process with varying $p$. The $\frac{1}{2}$-excess entropy $\mathbf{E}_{\frac{1}{2}}$ acts as a lower bound to all of them.}
\label{fig: perturbedcoin renyi entropies}
\end{figure}

The g-machine with smaller memory than the $\epsilon$-machine for this process is first found by Lohr \cite{lohr2009generative}. However, the advantage is only found for $p \in (0,\frac{1}{2})$ and it has 3 internal states. A more optimized g-machine is later found by Ruebeck-James-Mahoney-Crutchfield (RJMC) \cite{ruebeck2018prediction}, which is separated into two models for $0<p<\frac{1}{2}$ and $\frac{1}{2}< p < 1$; see Fig. \ref{fig: perturbed coin rjmc model}. The stationary distribution for the g-machines is given by
\begin{equation}
    \pi_{g} =\begin{cases} \begin{bmatrix}\frac{1-2p}{2-2p}\, , & \frac{1}{2-2p}\end{bmatrix}\, , \quad 0 < p < \frac{1}{2}\, , \\  \begin{bmatrix}\frac{1}{2p}\, , & \frac{2p-1}{2p} \end{bmatrix}\, , \quad \frac{1}{2} < p < 1\, ,\end{cases}
\end{equation}
and so the 2-statistical complexities read
\begin{equation}
    C_g^{(2)} = \begin{cases}
        -\log \frac{(1-2p)^2 + 1}{(2-2p)^2}\, ,\quad 0<p<\frac{1}{2}\, ,\\
        -\log \frac{1 + (2p-1)^2}{4p^2}\, ,\quad \frac{1}{2}<p<1\, .
    \end{cases}
\end{equation}
In Fig. \ref{fig: perturbedcoin renyi entropies}, we show the memory usage as quantified using notion of 2-statistical complexity for the different models of Perturbed Coin Process. As can be seen, the $\frac{1}{2}$-excess entropy provides a lower bound to all the 2-statistical complexity of the different models.

\begin{figure}[!htb]
    \centering
    \resizebox{1\linewidth}{!}{
    \begin{tikzpicture}[->, >=stealth', auto, semithick, node distance=3cm]
    \tikzstyle{every state}=[fill=white,draw=black,thick,text=black,scale=1]
    \node[draw=none, fill=none] (x) {};
    \node[state]    (A)[left of=x]    {$\tilde{\sigma}_{0,0}$};
    \node[state]    (B)[above of=x]   {$\tilde{\sigma}_{0,1}$};
    \node[state]    (C)[right of=x]    {$\tilde{\sigma}_{1,0}$};
    \path
    (A) edge[loop left] 	node{$0|1-p+q_1$}    (A)
        edge[above] 	node[right]{$0|-q_1$}    (B)
        edge[right]  node[above]{$1|p$}   (C)
    (B) edge[loop above]    node{$0|1-p+q_1$}    (B)
        edge[bend right, right]  node[left]{$0|-q_1$}   (A)
        edge[bend left, right]  node{$1|p$}     (C)
    (C) edge[loop right]    node{$1|1-p$}   (C)
        edge[bend left, left]   node[below]{$0|q_2$}  (A)
        edge[above]     node[left]{$0|p-q_2$}    (B);
    \end{tikzpicture}
    }
    \caption{n-machine representation of Perturbed Coin Process. New transition probabilities emerge from the state splitting, and we have parametrized these new degrees of freedom as $q_1,q_2\in \mathbb{R}$.}
    \label{fig: perturbed coin n-machine}
\end{figure}
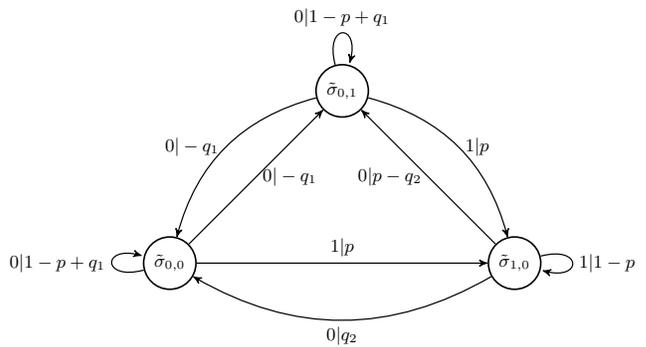

Now, we shall construct an n-machine that could yield memory advantage and even saturate the allowed lower limit. Starting from the $\epsilon$-machine model in Fig. \ref{fig: perturbed coin epsilon machine}, we consider splitting state $\sigma_0 \to \{\tilde{\sigma}_{0,0}, \tilde{\sigma}_{0,1}\}$ and keeping the state $\sigma_1\to \{\tilde{\sigma}_{1,0}\}$. We then introduce new free parameters $q_1,q_2$ in its state-to-state transition probabilities accordingly to Step 2 and draw out the model in Fig. \ref{fig: perturbed coin n-machine}. The stationary distribution of the n-machine is given by
\begin{equation}
\tilde{\pi} = \begin{bmatrix}
\frac{q_2-q_1}{2(p-2q_1)}, & \frac{p-q_2-q_1}{2(p-2q_1)}, & \frac{1}{2}
\end{bmatrix}\, ,
\end{equation}
where the stationary distributions correspond to the internal states $\tilde{\sigma}_{0,0}, \tilde{\sigma}_{0,1}, \tilde{\sigma}_{1,0}$ respectively. One can immediately see that indeed $\tilde{\pi}_{0,0}+\tilde{\pi}_{0,1} = \pi_{0}$ and $\tilde{\pi}_{1,0} = \pi_{1}$ in agreement with Eq. \eqref{eq: n-machine construct 3}. In order to have nonclassical memory advantage, i.e., $C_n^{(2)} < C_\mu^{(2)}$, it can be inferred easily that $\tilde{\pi}$ necessarily need to be a quasiprobability distribution. There are many ways in which one could make the n-machine here be the ideal model, that is, $C_n^{(2)} = \mathbf{E}_{\frac{1}{2}}$ is saturated. One such choice is to set
\begin{eqnarray}
q_1&=&0\, , \nonumber \\
q_2 &=& p \pm p^2\sqrt{\frac{1}{4p^2} + \frac{2\sqrt{1-p}}{p^{3/2}}}\, , \label{eq: perturbed coin n-machine parameters}
\end{eqnarray}
where both $\pm$ in $q_2$ does the job.

\begin{figure}[!htb]
    \centering
    \includegraphics[width=1.0\linewidth]{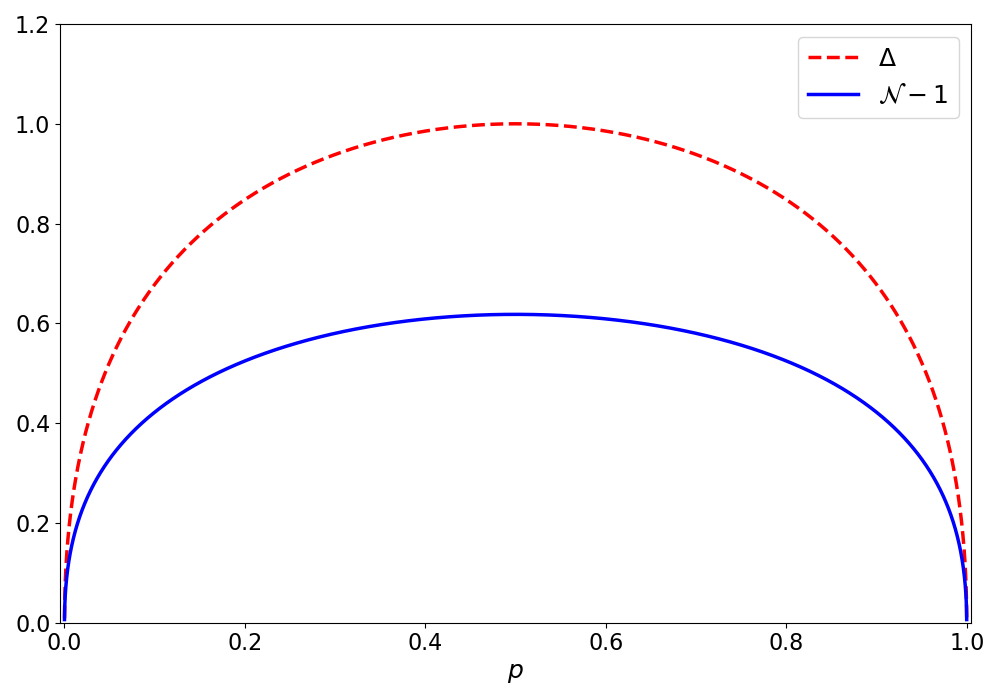}
    \caption{Negativity of the n-machine's stationary distribution for the Perturbed Coin Process. The negativity plotted is subtracted by a constant of 1 so that the lowest value of both quantities is the same, i.e., zero when no advantage is observed.}
    \label{fig: perturbed coin negativity}
\end{figure}

We can investigate how the negativity of the n-machine is related to the memory advantage. Consider the negativity of the n-machine defined as the $l_1$-norm of the quasiprobability stationary distribution:
\begin{equation}
    \mathcal{N} \equiv \norm{\tilde{\pi}}_1 = \sum_{k,l_k} |\tilde{\pi}_{k,l_k}| \geq 1
\end{equation}
with equality saturated iff $\tilde{\pi}$ is nonnegative. This definition of negativity has been proposed and used before; see \cite{veitch2014resource, oas2014exploring,pashayan2015estimating}. On the other hand, we use similar notion of memory advantage used in \cite{mahoney2016occam,ruebeck2018prediction} to quantify it as the relative difference
\begin{equation}
    \Delta = \frac{|C_n^{(2)} - C_\mu^{(2)}|}{C_\mu^{(2)}}\, .
\end{equation}
In Fig. \ref{fig: perturbed coin negativity}, we plot the negativity of n-machine with stationary distribution and parameters set in the above. One can check that as the negativity increases, the memory advantage also increases. This provides simple evidence on how negativity produces nonclassical advantage in this setting. Note that the same observations are obtained using mana, Eq. \eqref{eq: collision entropy simulation split terms}, as the measure of negativity.

\subsection{Simple Nonunifilar Source Process}

\begin{figure}[!htb]
\begin{subfigure}[t]{0.99\linewidth}
    \centering
    \begin{tikzpicture}[->, >=stealth', auto, semithick, node distance=3cm]
    \tikzstyle{every state}=[fill=white,draw=black,thick,text=black,scale=1]
    \node[state]    (A)                     {$\sigma_A$};
    \node[state]    (B)[right of=A]   		{$\sigma_B$};
    \path
    (A) edge[loop left] 		node{$0|p$}    (A)
        edge[bend left, above] 	node{$0|1-p$}      (B)
    (B) edge[loop right]  		node{$0|p$}    (B)
    	edge[bend left, below] 	node{$1|1-p$} 	 (A);
    \end{tikzpicture}
    \caption{}
    \label{fig: sns g-machine}
\end{subfigure}
\begin{subfigure}[t]{1.0\linewidth}
    \centering
    \begin{tikzpicture}[->, >=stealth', auto, semithick, node distance=2.5cm]
    \tikzstyle{every state}=[fill=white,draw=black,thick,text=black,scale=1]
    \node[state]    (A)                     {$\sigma_{1}$};
    \node[state]    (B)[right of=A]   		{$\sigma_{2}$};
    \node[state]    (K)[right of=B]   		{$\sigma_{n}$};
    \node[state]    (Z)[below of=B]   		{$\sigma_{0}$};
    \node[]     (F)[right of=K] {};
    \draw[] (B) edge[above] node{$0|\frac{\Phi(3)}{\Phi(2)}$} ($(B)!0.5!(K)$);
    \node at ($(B)!0.65!(K)$) {$\ldots$};
    \draw[] (K) edge[above] node{$0|\frac{\Phi(n+1)}{\Phi(n)}$}($(K)!0.6!(F)$);
    \node at ($(K)!0.70!(F)$) {$\ldots$};
    \path
    (A) edge[above] 	node[above]{$0|\frac{\Phi(2)}{\Phi(1)}$}      (B)
        edge[bend left, right] node{$1|\frac{\phi(1)}{\Phi(1)}$} (Z)
    (B) edge[bend left, right] node[pos=0.3]{$1|\frac{\phi(2)}{\Phi(2)}$} 	(Z)    
    (K) edge[bend left] 	node[pos=0.4]{$1|\frac{\phi(n)}{\Phi(n)}$} (Z)
    (Z) edge[bend left, left] node{$0|1$} 	(A);
    \end{tikzpicture}
    \caption{}
    \label{fig: SNS epsilon machine}
\end{subfigure}
\caption{(a) g-machine and (b) $\epsilon$-machine representation of the SNS Process.}
\label{fig: sns hmm representation}
\end{figure}
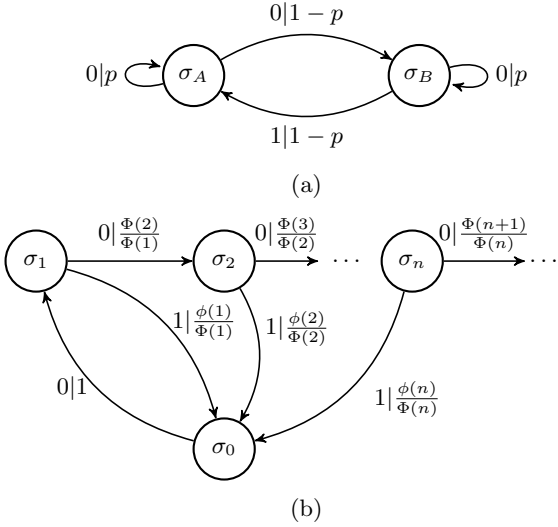

In the next example, we study a more complicated process known as the Simple Nonunifilar Source (SNS) Process, which is a type of (discrete-time) renewal process. This process has been studied extensively in \cite{marzen2015informational,elliott2021memory}, and so they provide a useful resource for our calculations. In the following, we will state the quantities relevant to us without delving too deeply into them. In particular, we have the {\it }waiting-time distribution $\phi(n) = np^{n-1}(1-p)^2, \sum_{n=0}^\infty \phi(n) = 1$, which is the probability of seeing $n$ consecutive 0's in between a 1; the {\it surviving probability} $\Phi(n) \equiv \sum_{k=n}^\infty \phi(k)$; and the {\it mean firing rate} $\mu = 1/\sum_{n=0}^\infty \Phi(n)$. With these in mind, the g-machine and $\epsilon$-machine representation of the SNS Process is shown in Fig. \ref{fig: sns hmm representation}. It can be noticed immediately that the generative model is fairly simple with only two internal states, while the $\epsilon$-machine has an infinite number of causal states. The g-machine has a simple stationary distribution of $\pi = [\frac{1}{2}, \frac{1}{2}]$ for all $p$ while for the $\epsilon$-machine, the causal state $\sigma_n$ has distribution $\pi_n = \mu \Phi(n)$. 

\begin{figure}[!htb]
    \centering
    \includegraphics[width=1.0\linewidth]{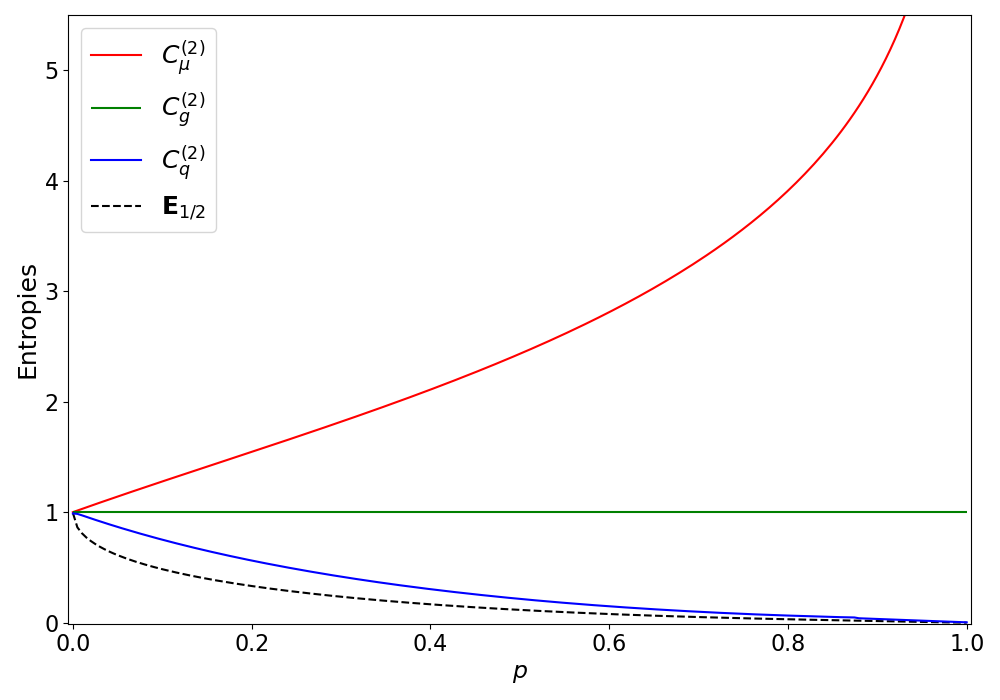}
    \caption{Comparison of the memory usage by the different models for the SNS Process with varying $p$.}
    \label{fig: sns renyi entropies}
\end{figure}

For its q-machine, it has been established before that the states are encoded as
\begin{equation}
    \ket{\sigma_n} = \sum_{k=0}^\infty \sqrt{\frac{\Phi(n+k)}{\Phi(n)}}\ket{k}\, .
\end{equation}
The details to calculate $\mathbf{E}_{\frac{1}{2}}$ are shown in Appendix \ref{sec: appendix examples details}. We plot the memory usage of the different models in Fig. \ref{fig: sns renyi entropies}.

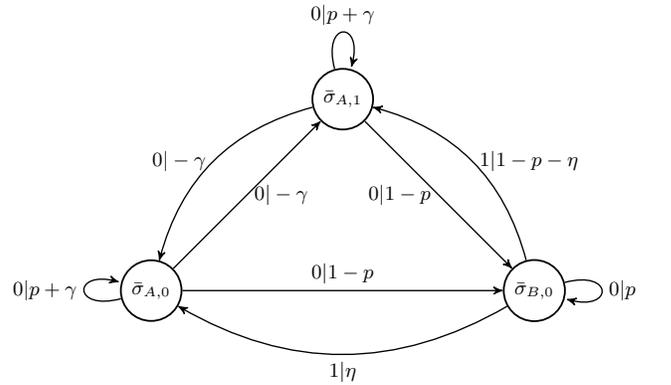
\begin{figure}[!htb]
    \centering
    \resizebox{1\linewidth}{!}{
    \begin{tikzpicture}[->, >=stealth', auto, semithick, node distance=3cm]
    \tikzstyle{every state}=[fill=white,draw=black,thick,text=black,scale=1]
    \node[draw=none, fill=none] (x) {};
    \node[state]    (A)[left of=x]    {$\bar{\sigma}_{A,0}$};
    \node[state]    (B)[above of=x]   {$\bar{\sigma}_{A,1}$};
    \node[state]    (C)[right of=x]    {$\bar{\sigma}_{B,0}$};
    \path
    (A) edge[loop left] 	node{$0|p+\gamma$}    (A)
        edge[above] 	node[right]{$0|-\gamma$}    (B)
        edge[right]  node[above]{$0|1-p$}   (C)
    (B) edge[loop above]    node{$0|p+\gamma$}    (B)
        edge[bend right, right]  node[left]{$0|-\gamma$}   (A)
        edge[right]  node[left]{$0|1-p$}     (C)
    (C) edge[loop right]    node{$0|p$}   (C)
        edge[bend left, left]   node[below]{$1|\eta$}  (A)
        edge[bend right]     node[right]{$1|1-p-\eta$}    (B);
    \end{tikzpicture}
    }
    \caption{n-machine representation of SNS Process parameterized by $\gamma,\eta\in \mathbb{R}$.}
    \label{fig: sns n-machine}
\end{figure}

\begin{figure}[!htb]
    \centering
    \includegraphics[width=1.0\linewidth]{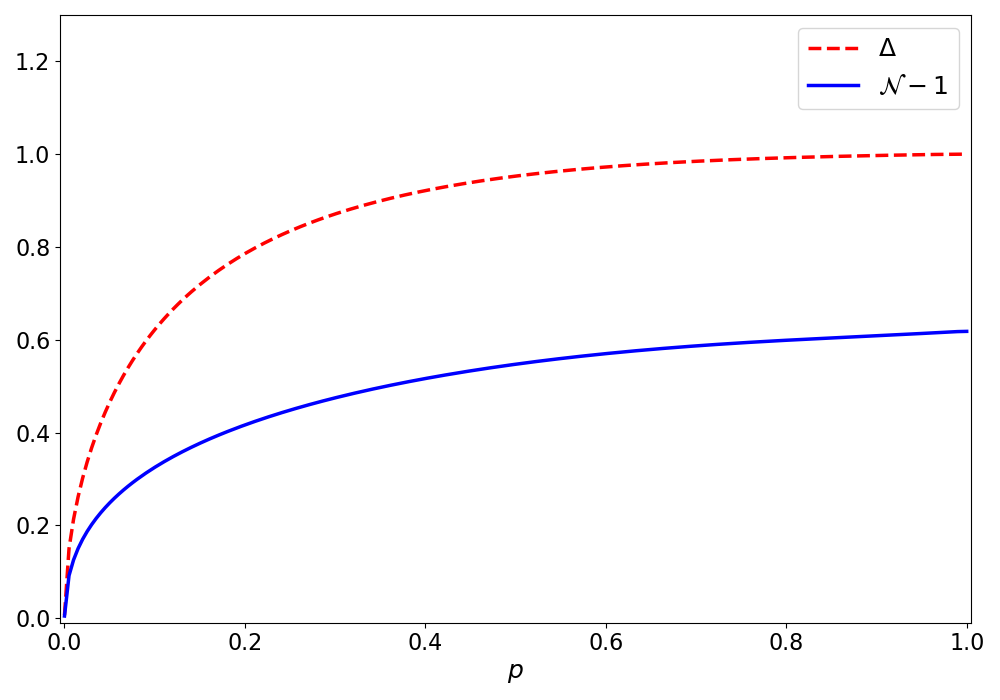}
    \caption{Comparison between negativity and memory advantage for SNS Process.}
    \label{fig: sns negativity}
\end{figure}

Now, we will construct the n-machine for the SNS Process. As mentioned in the previous section, the protocol to construct the n-machine can also start from any classical generative model, and not necessarily the $\epsilon$-machine. As such, we consider starting from the g-machine as seen in Fig. \ref{fig: sns g-machine} and construct the n-machine as drawn out in Fig. \ref{fig: sns n-machine}. We find the stationary distribution of the n-machine to be 
\begin{equation}
\tilde{\pi} =\frac{1}{2} \begin{bmatrix}
\frac{\gamma-\eta}{2\gamma+p-1}, & \frac{p+\gamma+\eta-1}{2\gamma+p-1}, & 1
\end{bmatrix}
\end{equation}
corresponding to the internal states $\tilde{\sigma}_{A,0}, \tilde{\sigma}_{A,1}, \tilde{\sigma}_{B,0}$. To saturate $C_n^{(2)} = \mathbf{E}_{\frac{1}{2}}$, we could then take, for instance, 
\begin{eqnarray}
\gamma &=& 0\, , \nonumber \\
\eta &=& \frac{1}{2}\left[1-p\pm (p-1)\sqrt{-3+8M}\right]\, , \label{eq: sns process n-machine parameters}
\end{eqnarray}
where 
\begin{equation}\label{eq: quantity inside log excess entropy sns process}
    M = \sum_{m=0}^\infty \left(\sum_{n=0}^\infty \mu \sqrt{ \phi(m+n)\Phi(n)} \right)^2\, .
\end{equation}
Similar to the previous case, we plot the negativity of the n-machine's state distribution and compare it with the memory advantage gained in Fig. \ref{fig: sns negativity}. Again, we can see the profile between the negativity and memory advantage takes a similar form. This exemplifies the relationships between negativity and memory advantage. 

\section{Discussions}

\subsection{Additional remarks about n-machine}

\begin{figure}[!htb]
\centering
\begin{subfigure}[t]{0.99\linewidth}
\centering
\begin{tikzpicture}[->, >=stealth', auto, semithick, node distance=3cm]
\tikzstyle{every state}=[fill=white,draw=black,thick,text=black,scale=1]
\node[state]    (A)                     {$\sigma_0$};
\node[state]    (B)[right of=A]   		{$\sigma_1$};
\path
(A) edge[loop left] 		node{$0|p$}    (A)
    edge[bend left, above] 	node{$1|1-p$}    (B)
(B)	edge[bend left, below] 	node{$0|1$} 	 (A);
\end{tikzpicture}
\caption{}
\label{fig: golden mean hmm}
\end{subfigure}
\begin{subfigure}[t]{0.99\linewidth}
\raggedright
    \begin{tikzpicture}[->, >=stealth', auto, semithick, node distance=2cm]
    \tikzstyle{every state}=[fill=white,draw=black,thick,text=black,scale=1]
    \node[draw=none, fill=none] (x) {};
    \node[state]    (A)[left of=x]    {$\tilde{\sigma}_{0,0}$};
    \node[state]    (B)[above of=x]   {$\tilde{\sigma}_{0,1}$};
    \node[state]    (C)[right of=x]    {$\tilde{\sigma}_{1,0}$};
    \path
    (A) edge[loop left] 	node{$0|p+q$}    (A)
        edge[above] 	node[right]{$0|-q_1$}    (B)
        edge[right]  node[above]{$1|1-p$}   (C)
    (B) edge[loop above]    node{$0|p+q$}    (B)
        edge[bend right, right]  node[left]{$0|-q$}   (A)
        edge[bend left, right]  node{$1|1-p$}     (C)
    (C) edge[bend left, left]   node[below]{$0|\frac{1}{2}$}  (A)
        edge[above]     node[left]{$0|\frac{1}{2}$}    (B);
    \end{tikzpicture}
    \caption{}
    \label{fig: golden mean bad n-machine}
\end{subfigure}
\caption{(a) $\epsilon$-machine of Golden Mean Process. (b) Example of `bad' n-machine for Golden Mean Process exhibiting negative transition probabilities but no memory advantage.}
\end{figure}
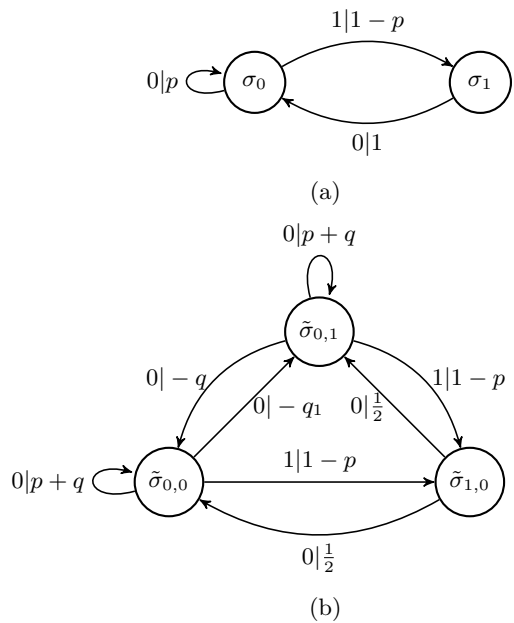

Although negativity appears to be a necessary resource for nonclassical advantage, it is not sufficient on its own. To illustrate this, let us use the Golden Mean Process \cite{mahoney2009information}, whose $\epsilon$-machine representation is depicted in Fig. \ref{fig: golden mean hmm}. Its stationary distribution is $\pi = [\frac{1}{2-p}, \frac{1-p}{2-p}]$ so $C_\mu^{(2)} = -\log\left[(2-2p+p^2)/(p-2)^2\right]$. Now, consider that we construct the n-machine as seen in Fig. \ref{fig: golden mean bad n-machine} with $q\in\mathbb{R}$ as the new degree of freedom. This n-machine has a stationary distribution 
\begin{equation}
    \tilde{\pi} = \frac{1-p}{2-p}\begin{bmatrix}\frac{1}{2-2p}\, , & \frac{1}{2-2p}\, ,& 1\end{bmatrix}
\end{equation}
From here, it is clear that $\tilde{\pi}$ is independent of the new free parameter $q$ and that $C_n^{(2)} > C_\mu^{(2)}$ everywhere. Thus, no advantage can be gained through this construction, however much we tune $q$. This tells us that although negativity seems to be necessary, it is not sufficient --- this is congruent with previous findings in \cite{veitch2012negative}. The n-machine constructed here can be thought of as a `bad' n-machine and to obtain a better one with memory advantage, this can be done by considering setting $\tilde{T}^{(0)}_{1,0; 0,0} = q'$, $\tilde{T}^{(0)}_{1,0;0,1} = 1-q'$ and taking $q'<0$. 

We remark that although the n-machine constructed here is an instance of an ideal generative model, there are many other ways to construct an ideal model. Given a model, one could find another model that generates the same stochastic process by using an invertible linear map $Z$ (as noted in \cite{fanizza2024quantum}). This can be seen by considering two generators $g_1\equiv(\mathcal{A}, \bm{\mathcal{S}}, \{T^{(x)}\}, \pi)$ and $g_2 \equiv (\mathcal{A}, \bm{\mathcal{S}}', \{T'^{(x)}\}, \pi')$ which are related by an invertible linear map $Z$ that satisfies $\pi' = \pi Z^{-1}$, $T'^{(x)} = ZT^{(x)}Z^{-1}$, and $\bm{1}=Z\bm{1}$. This general mapping allows one to transform a classical model into another model with a desirable property, such as lower memory complexity. For instance, considering the Perturbed Coin process again, one can find a model that has negative transition probabilities but a non-negative stationary distribution\footnote{This is related to the generalization of the Perron-Frobenius theorem \cite{perron1907theorie,frobenius1912matrizen}, which appears to be extremely difficult to prove generally but several cases has been shown to exist \cite{tarazaga2001perron,johnson2004matrices,noutsos2006perron,curgus2015somewhat,chruscinski2015pseudo}.} (see Appendix \ref{sec: appedix gpt hmm linear map}). Although the model we found using this map appears minimal, it is difficult to check whether it satisfies the data processing inequality, as mutual information may not be well-defined for the obtained transition probabilities --- unlike in the case of n-machine. Moreover, there remains a difficulty in characterizing an optimal invertible linear transformation.

Returning to our n-machine protocol, there are still many improvements that can be made and many more features to be studied in the future. For instance, the construction here necessarily increases the number of internal states used by the n-machine as compared to its previous starting point $\epsilon$-machine. In other words, $C_n^0 > C_\mu^0$. In some tasks, however, reducing the number of states takes priority, as considered in \cite{elliott2020extreme}. Another point of improvement is to minimize the negativity, as we have seen in the `bad' n-machine example above --- inserting negativity is not always beneficial. Thus, one should consider a construction that injects negativity in an effective manner. This could be done by appending Step 3 of the protocol with a negativity minimization task. We note, however, that the optimization problem over negativity measure is a nonlinear problem and could be challenging to solve with the current framework \cite{oas2014exploring}.

Although negativity appears to be a feature of the nonclassical memory reduction, it is unclear how it improves the model's efficiency. For g-machine, it appears that obtaining nonzero {\it oracular information}, i.e., $\zeta = I[\mathcal{S};\Future|\Past] \geq 0$, is beneficial in reducing the model's memory. Although $\zeta$ still contributes to the overall memory requirement, the model's {\it crypticity} $\chi = I[\Past;\mathcal{S}|\Future]$ reduces significantly more resulting to an overall decrease in memory.\footnote{There is also an additional information component contributing to the statistical complexity known as {\it gauge information} $\varphi = H[\mathcal{S}|\Past,\Future]$ but is found to be less significant than the crypticity and oracular information.} Here, we conjecture that the n-machine has somehow gained oracular information and reduced its crypticity. Although it has been shown above that $I_\frac{1}{2}[\tilde{\mathcal{S}};\Future] = \mathbf{E}_{\frac{1}{2}}$ indicating that there is no oracular information, the R{\'e}nyi-$\alpha$ entropies does not satisfy a simple information diagram \cite{crutchfield2010synchronization} relationship that affords us this implication. One would need to define and study a R{\'e}nyi-$\alpha$ generalization of the conditional mutual information that extends naturally from Eq. \eqref{eq: hv alpha mutual information}. We shall leave this as future work.

Throughout the paper and as part of the main problem we consider, we have focused on finding models with memory obeying the inequality $C_n^{(2)} \geq \mathbf{E}_{\frac{1}{2}}$. This arises as a mathematical limitation that naturally emerged in the classical and quantum regimes and thus imposed for GPTs too. However, in the post-quantum domain, it is easy to find a model that could violate such inequality if there is too much `negativity'. For instance, adding a small value $\alpha > 0$ to the parameters in Eq. \eqref{eq: perturbed coin n-machine parameters} or \eqref{eq: sns process n-machine parameters} will violate the inequality. As a matter of fact, one can go beyond to reach zero or negative values for $H_2[\tilde{\mathcal{S}}]$ if the negativity is unrestricted. At this point, little is understood about this issue, including whether there is a more physically motivated explanation for the inequality that accounts for post-quantum models, or perhaps a modification to their definition is required. We hope that this opens up an interesting discussion for the community. 

\subsection{Preliminary comparison with q-machine in quasiprobability representation}

Aside from obtaining realizations of the ideal model, we believe that the n-machine will be potentially useful in providing us with a further understanding when compared with the q-machine. Although seemingly different in construction, we can utilize \textit{quasiprobability representation} \cite{ferrie2008frame,ferrie2009framed,ferrie2011quasi} of the q-machine, or HQMM generally, to enable direct comparison.
Under this representation, it puts the HQMM framework, typically described in terms of operators in the Hilbert space, into vectors or matrices with real elements --- on equal footing with the classical and GPT HMMs here. For instance, a quantum state described by a density operator will be represented as a quasiprobability distribution, and channel operations will be described as quasi-stochastic matrices. As such, direct comparison between classical, quantum, and post-quantum models can be made formally. Additionally, it will enable us to visualize the q-machine as a directed graph, as the $\epsilon$-machine and n-machine have been represented here, albeit with negative transition probabilities. 

We show an explicit example in Appendix \ref{sec: appendix qpr} for the q-machine of Perturbed Coin Process under the {\it discrete Wigner representation} \cite{wootters1987wigner} and discussed its features. This is particularly interesting when compared with the n-machine's properties. Although the q-machine does not satisfy all the properties of the n-machine, some of the basic ideas, such as the state splitting (Step 1) and negativity injection in transition probabilities (Step 2), can be found. Therefore, we believe that the proposed n-machine construction can be valuable for further understanding the source of non-classical advantage achieved by quantum models. The ultimate goal of this study is then to find a principle that would immediately characterize physical theories in the likes of Ref. \cite{pawlowski2009information}. It would also be interesting to show if q-machines can be represented in a way that satisfies all n-machine's properties or vice versa.

\section{Conclusion}

Studies of classical and quantum hidden Markov models (HMMs), like $\epsilon$-machine and q-machine, are generally not ideal models -- models whose memory, as quantified by its statistical complexity, is equal to the excess entropy. Recognizing that classical and quantum may not be sufficient, we investigate post-quantum theories within generalized probabilistic theories (GPT) of HMM and propose an intuitive construction of an ideal model that we refer to as negative-machine (n-machine). In particular, we show that ideal models can be constructed by splitting the internal states of a classical machine and injecting negative probabilities (quasiprobabilities) into its transition matrices, without sacrificing statistical fidelity. We have also proposed the use of R{\'e}nyi-2 entropy as a measure of memory (statistical complexity) that can be applied for classical and non-classical models, while rigorously showed that many of the known properties and relationships still hold.

We demonstrate instances of n-machine for paradigmatic stochastic processes, namely the Perturbed Coin Process and Single Nonunifilar Source Process. Through these examples, we found that measures of negativity, such as the sum negativity, have a close relationship with the nonclassical memory advantage as quantified by the relative difference between the classical and n-machine's memory. As a result, we believe that negativity has an important role in achieving nonclassical advantage in the setting of stochastic process modeling, similar to negativity's connection to other nonclassical features previously studied \cite{abramsky2014operational, onggadinata2023simulations, spekkens2008negativity, booth2022contextuality, veitch2012negative, howard2014contextuality, pashayan2015estimating, kaszlikowski2021little}. It would also be worthwhile to study the connection between negativity and contextuality as a resource for nonclassical memory advantage \cite{cabello2018optimal}.

n-machine enables the study of stochastic process modeling beyond classical and quantum, which we believe can be helpful in understanding how non-classical models achieve memory compression. Preliminary comparison with the q-machine in quasiprobability representation also reveals some insights that the steps taken to construct n-machine also appear in the q-machine, although a deeper investigation should be conducted. It would also be interesting to extend the findings in \cite{aghamohammadi2017ambiguity}, which looks into the ``simplicity" of two stochastic processes modeled classically or quantally. Here, they considered the relative simplicity of two stochastic processes, A and B, and assumed that A is simpler than B classically. The paper showed that the quantum model of A and B can either preserve (A simpler than B) or reverse (B simpler than A) this relationship, which is referred to as consistent or ambiguous, respectively. Hence, it would be interesting to extend this study to show if the consistency and ambiguity of the processes are generally applicable for classical-to-nonclassical (quantum and post-quantum) or if a transition in behavior can also occur between quantum and post-quantum regimes. 

Many more works can be done to improve the ideal machine construction and explore its features, but we believe that this work contributes to the study of general theories of stochastic process modeling and could be relevant in finding physical principles in this setting. 
Moreover, it would be interesting to explore our approach in stochastic simulation of open quantum systems \cite{breuer1995stochastic}, non-Markovian phenomena \cite{milz2021quantum} and thermodynamics \cite{garner2017thermodynamics, elliott2021memory}. In the latter, it has been shown that memory-efficient models exhibit the least heat dissipation, with quantum models demonstrated to be more thermally efficient than classical models. Thus, it would perhaps be of interest to study and show if ideal models are more thermally efficient than quantum or if a thermodynamic principle has been violated.

We also note that the q-machine can be modeled in the quantum combs or process tensor formalism \cite{chiribella2009theoretical,milz2021quantum}, or more generally as a process matrix~\cite{oreshkov2012quantum}.
In this formalism, the entire quantum model (as illustrated in Fig.~\ref{fig: quantum_HMM}) outputting a finite sequence of symbols can be represented as a single black-box object, for which the experimenter obtains an output $|x_t\rangle$ at each timestep and inputs a $|0\rangle$ back to it.
It would be interesting to analyze in this framework whether it is possible to construct a model that saturates the excess entropy for a certain stochastic process, i.e. given a certain causality constraint, and how it relates to the amount of negativity of the model.
On the other hand, it is also an interesting future direction to analyze in these frameworks how negativity relates to the size of the memory register necessary to generate a certain stochastic process.

\section*{Acknowledgments}
This work is supported by the National Research Foundation, Singapore through the National Quantum Office, hosted in A*STAR, under its Centre for Quantum Technologies Funding Initiative (S24Q2d0009), and the Singapore Ministry of Education Tier 1 Grant RT4/23. KO was supported by the CQT PhD Scholarship. AT is supported by the CQT PhD Scholarship, the Google PhD Fellowship program, and CQT Young Researcher Career Development Grant. We thank Clive Cenxin Aw for discussions. 

\normalem

\bibliographystyle{quantum}
\bibliography{references}

\begin{thebibliography}{100}

\bibitem{rabiner1986introduction}
L.~Rabiner and B.~Juang.
\newblock ``An introduction to hidden {M}arkov models''.
\newblock \href{https://dx.doi.org/10.1109/MASSP.1986.1165342}{IEEE ASSP Magazine {\bf 3}, 4--16}~(1986).

\bibitem{vidyasagar2014hidden}
M.~Vidyasagar.
\newblock ``Hidden {M}arkov processes: Theory and applications to biology''.
\newblock \href{https://dx.doi.org/https://doi.org/10.2307/j.ctt6wq0db}{Princeton University Press}. ~(2014).

\bibitem{jelinek1998statistical}
Frederick Jelinek.
\newblock ``Statistical methods for speech recognition''.
\newblock MIT press. ~(1998).
\newblock  url:~\url{https://mitpress.mit.edu/9780262100663/statistical-methods-for-speech-recognition/}.

\bibitem{crutchfield1997statistical}
James~P. Crutchfield and David~P. Feldman.
\newblock ``Statistical complexity of simple one-dimensional spin systems''.
\newblock \href{https://dx.doi.org/10.1103/PhysRevE.55.R1239}{Phys. Rev. E {\bf 55}, R1239--R1242}~(1997).

\bibitem{suen2017classical}
Whei~Yeap Suen, Jayne Thompson, Andrew~JP Garner, Vlatko Vedral, and Mile Gu.
\newblock ``The classical-quantum divergence of complexity in modelling spin chains''.
\newblock \href{https://dx.doi.org/10.22331/q-2017-08-11-25}{Quantum {\bf 1}, 25}~(2017).

\bibitem{ghahramani1995factorial}
Zoubin Ghahramani and Michael~I. Jordan.
\newblock ``Factorial hidden {M}arkov models''.
\newblock In Proceedings of the 9th International Conference on Neural Information Processing Systems.
\newblock Page 472–478.
\newblock NIPS'95Cambridge, MA, USA~(1995). MIT Press.
\newblock  url:~\url{https://dl.acm.org/doi/10.5555/2998828.2998895}.

\bibitem{fine1998hierarchical}
Shai Fine, Yoram Singer, and Naftali Tishby.
\newblock ``The hierarchical hidden {M}arkov model: Analysis and applications''.
\newblock \href{https://dx.doi.org/10.1023/A:1007469218079}{Machine learning {\bf 32}, 41--62}~(1998).

\bibitem{haslinger2010computational}
Robert Haslinger, Kristina~Lisa Klinkner, and Cosma~Rohilla Shalizi.
\newblock ``The computational structure of spike trains''.
\newblock \href{https://dx.doi.org/10.1162/neco.2009.12-07-678}{Neural Computation {\bf 22}, 121--157}~(2010).

\bibitem{yang2008increasing}
Jae-Suk Yang, Wooseop Kwak, Taisei Kaizoji, and In-mook Kim.
\newblock ``Increasing market efficiency in the stock markets''.
\newblock \href{https://dx.doi.org/10.1140/epjb/e2008-00050-0}{The European Physical Journal B {\bf 61}, 241--246}~(2008).

\bibitem{crutchfield2010synchronization}
James~P. Crutchfield, Christopher~J. Ellison, Ryan~G. James, and John~R. Mahoney.
\newblock ``{Synchronization and control in intrinsic and designed computation: An information-theoretic analysis of competing models of stochastic computation}''.
\newblock \href{https://dx.doi.org/10.1063/1.3489888}{Chaos: An Interdisciplinary Journal of Nonlinear Science {\bf 20}, 037105}~(2010).

\bibitem{crutchfield1989inferring}
James~P. Crutchfield and Karl Young.
\newblock ``Inferring statistical complexity''.
\newblock \href{https://dx.doi.org/10.1103/PhysRevLett.63.105}{Phys. Rev. Lett. {\bf 63}, 105--108}~(1989).

\bibitem{shalizi2001computational}
Cosma~Rohilla Shalizi and James~P Crutchfield.
\newblock ``Computational mechanics: Pattern and prediction, structure and simplicity''.
\newblock \href{https://dx.doi.org/10.1023/A:1010388907793}{Journal of statistical physics {\bf 104}, 817--879}~(2001).

\bibitem{lohr2009generative}
Wolfgang L\"{o}hr and Nihat Ay.
\newblock ``On the generative nature of prediction''.
\newblock \href{https://dx.doi.org/10.1142/S0219525909002143}{Advances in Complex Systems {\bf 12}, 169--194}~(2009).

\bibitem{lohr2009non}
Wolfgang L{\"o}hr and Nihat Ay.
\newblock ``Non-sufficient memories that are sufficient for prediction''.
\newblock In Complex Sciences.
\newblock Pages 265--276.
\newblock Berlin, Heidelberg~(2009). Springer Berlin Heidelberg.
\newblock  url:~\url{https://doi.org/10.1007/978-3-642-02466-5_25}.

\bibitem{lohr2012predictive}
Wolfgang L{\"o}hr.
\newblock ``Predictive models and generative complexity''.
\newblock \href{https://dx.doi.org/10.1007/s11424-012-9173-x}{Journal of Systems Science and Complexity {\bf 25}, 30--45}~(2012).

\bibitem{ruebeck2018prediction}
Joshua~B. Ruebeck, Ryan~G. James, John~R. Mahoney, and James~P. Crutchfield.
\newblock ``Prediction and generation of binary {M}arkov processes: Can a finite-state fox catch a {M}arkov mouse?''.
\newblock \href{https://dx.doi.org/10.1063/1.5003041}{Chaos: An Interdisciplinary Journal of Nonlinear Science {\bf 28}, 013109}~(2018).

\bibitem{thompson2018causal}
Jayne Thompson, Andrew J.~P. Garner, John~R. Mahoney, James~P. Crutchfield, Vlatko Vedral, and Mile Gu.
\newblock ``Causal asymmetry in a quantum world''.
\newblock \href{https://dx.doi.org/10.1103/PhysRevX.8.031013}{Phys. Rev. X {\bf 8}, 031013}~(2018).

\bibitem{gu2012quantum}
Mile Gu, Karoline Wiesner, Elisabeth Rieper, and Vlatko Vedral.
\newblock ``Quantum mechanics can reduce the complexity of classical models''.
\newblock \href{https://dx.doi.org/10.1038/ncomms1761}{Nature communications {\bf 3}, 762}~(2012).

\bibitem{mahoney2016occam}
John~R Mahoney, Cina Aghamohammadi, and James~P Crutchfield.
\newblock ``Occam’s quantum strop: Synchronizing and compressing classical cryptic processes via a quantum channel''.
\newblock \href{https://dx.doi.org/10.1038/srep20495}{Scientific reports {\bf 6}, 20495}~(2016).

\bibitem{riechers2016minimized}
Paul~M. Riechers, John~R. Mahoney, Cina Aghamohammadi, and James~P. Crutchfield.
\newblock ``Minimized state complexity of quantum-encoded cryptic processes''.
\newblock \href{https://dx.doi.org/10.1103/PhysRevA.93.052317}{Phys. Rev. A {\bf 93}, 052317}~(2016).

\bibitem{binder2018practical}
Felix~C. Binder, Jayne Thompson, and Mile Gu.
\newblock ``Practical unitary simulator for non-{M}arkovian complex processes''.
\newblock \href{https://dx.doi.org/10.1103/PhysRevLett.120.240502}{Phys. Rev. Lett. {\bf 120}, 240502}~(2018).

\bibitem{liu2019optimal}
Qing Liu, Thomas~J. Elliott, Felix~C. Binder, Carlo Di~Franco, and Mile Gu.
\newblock ``Optimal stochastic modeling with unitary quantum dynamics''.
\newblock \href{https://dx.doi.org/10.1103/PhysRevA.99.062110}{Phys. Rev. A {\bf 99}, 062110}~(2019).

\bibitem{elliott2020extreme}
Thomas~J. Elliott, Chengran Yang, Felix~C. Binder, Andrew J.~P. Garner, Jayne Thompson, and Mile Gu.
\newblock ``Extreme dimensionality reduction with quantum modeling''.
\newblock \href{https://dx.doi.org/10.1103/PhysRevLett.125.260501}{Phys. Rev. Lett. {\bf 125}, 260501}~(2020).

\bibitem{elliott2022quantum}
Thomas~J. Elliott, Mile Gu, Andrew J.~P. Garner, and Jayne Thompson.
\newblock ``Quantum adaptive agents with efficient long-term memories''.
\newblock \href{https://dx.doi.org/10.1103/PhysRevX.12.011007}{Phys. Rev. X {\bf 12}, 011007}~(2022).

\bibitem{wiesner2008computation}
Karoline Wiesner and James~P. Crutchfield.
\newblock ``Computation in finitary stochastic and quantum processes''.
\newblock \href{https://dx.doi.org/https://doi.org/10.1016/j.physd.2008.01.021}{Physica D: Nonlinear Phenomena {\bf 237}, 1173--1195}~(2008).

\bibitem{monras2012hidden}
Alex Monras, Almut Beige, and Karoline Wiesner.
\newblock ``Hidden quantum markov models and non-adaptive read-out of many-body states''~(2012).
\newblock  \href{http://arxiv.org/abs/1002.2337}{arXiv:1002.2337}.

\bibitem{monras2016quantum}
Alex Monràs and Andreas Winter.
\newblock ``Quantum learning of classical stochastic processes: The completely positive realization problem''.
\newblock \href{https://dx.doi.org/10.1063/1.4936935}{Journal of Mathematical Physics {\bf 57}, 015219}~(2016).

\bibitem{loomis2020thermal}
Samuel~P. Loomis and James~P. Crutchfield.
\newblock ``Thermal efficiency of quantum memory compression''.
\newblock \href{https://dx.doi.org/10.1103/PhysRevLett.125.020601}{Phys. Rev. Lett. {\bf 125}, 020601}~(2020).

\bibitem{ferrie2010necessity}
Christopher Ferrie, Ryan Morris, and Joseph Emerson.
\newblock ``Necessity of negativity in quantum theory''.
\newblock \href{https://dx.doi.org/10.1103/PhysRevA.82.044103}{Phys. Rev. A {\bf 82}, 044103}~(2010).

\bibitem{abramsky2014operational}
Samson Abramsky and Adam Brandenburger.
\newblock ``An operational interpretation of negative probabilities and no-signalling models''.
\newblock Pages 59--75.
\newblock Springer International Publishing. Cham~(2014).
\newblock  url:~\url{https://doi.org/10.1007/978-3-319-06880-0_3}.

\bibitem{onggadinata2023simulations}
Kelvin Onggadinata, Pawel Kurzynski, and Dagomir Kaszlikowski.
\newblock ``Simulations of quantum nonlocality with local negative bits''.
\newblock \href{https://dx.doi.org/10.1103/PhysRevA.108.032204}{Phys. Rev. A {\bf 108}, 032204}~(2023).

\bibitem{spekkens2008negativity}
Robert~W. Spekkens.
\newblock ``Negativity and contextuality are equivalent notions of nonclassicality''.
\newblock \href{https://dx.doi.org/10.1103/PhysRevLett.101.020401}{Phys. Rev. Lett. {\bf 101}, 020401}~(2008).

\bibitem{booth2022contextuality}
Robert~I. Booth, Ulysse Chabaud, and Pierre-Emmanuel Emeriau.
\newblock ``Contextuality and {W}igner negativity are equivalent for continuous-variable quantum measurements''.
\newblock \href{https://dx.doi.org/10.1103/PhysRevLett.129.230401}{Phys. Rev. Lett. {\bf 129}, 230401}~(2022).

\bibitem{veitch2012negative}
Victor Veitch, Christopher Ferrie, David Gross, and Joseph Emerson.
\newblock ``Negative quasi-probability as a resource for quantum computation''.
\newblock \href{https://dx.doi.org/10.1088/1367-2630/14/11/113011}{New Journal of Physics {\bf 14}, 113011}~(2012).

\bibitem{howard2014contextuality}
Mark Howard, Joel Wallman, Victor Veitch, and Joseph Emerson.
\newblock ``Contextuality supplies the ‘magic’ for quantum computation''.
\newblock \href{https://dx.doi.org/https://doi.org/10.1038/nature13460}{Nature {\bf 510}, 351--355}~(2014).

\bibitem{pashayan2015estimating}
Hakop Pashayan, Joel~J. Wallman, and Stephen~D. Bartlett.
\newblock ``Estimating outcome probabilities of quantum circuits using quasiprobabilities''.
\newblock \href{https://dx.doi.org/10.1103/PhysRevLett.115.070501}{Phys. Rev. Lett. {\bf 115}, 070501}~(2015).

\bibitem{kaszlikowski2021little}
Dagomir Kaszlikowski and Pawe{\l} Kurzy{\'n}ski.
\newblock ``A little bit of classical magic to achieve (super-) quantum speedup''.
\newblock \href{https://dx.doi.org/10.1007/s10701-021-00461-w}{Foundations of Physics {\bf 51}, 55}~(2021).

\bibitem{barrett2008information}
Jonathan Barrett.
\newblock ``Information processing in generalized probabilistic theories''.
\newblock \href{https://dx.doi.org/10.1103/PhysRevA.75.032304}{Phys. Rev. A {\bf 75}, 032304}~(2007).

\bibitem{fanizza2024quantum}
Marco Fanizza, Josep Lumbreras, and Andreas Winter.
\newblock ``Quantum theory in finite dimension cannot explain every general process with finite memory''.
\newblock \href{https://dx.doi.org/https://doi.org/10.1007/s00220-023-04913-4}{Communications in Mathematical Physics {\bf 405}, 1--24}~(2024).

\bibitem{dong2023promotion}
Anqi Dong, Tryphon~T. Georgiou, and Allen Tannenbaum.
\newblock ``Promotion/inhibition effects in networks: A model with negative probabilities''~(2023).
\newblock  \href{http://arxiv.org/abs/2307.07738}{arXiv:2307.07738}.

\bibitem{crutchfield1994observing}
James~P. Crutchfield.
\newblock ``Observing complexity and the complexity of observation''.
\newblock In Harald Atmanspacher and Gerhard~J. Dalenoort, editors, Inside Versus Outside.
\newblock Pages 235--272.
\newblock Berlin, Heidelberg~(1994). Springer Berlin Heidelberg.
\newblock  url:~\url{https://doi.org/10.1007/978-3-642-48647-0_14}.

\bibitem{ellison2011information}
Christopher~J. Ellison, John~R. Mahoney, Ryan~G. James, James~P. Crutchfield, and Jörg Reichardt.
\newblock ``Information symmetries in irreversible processes''.
\newblock \href{https://dx.doi.org/10.1063/1.3637490}{Chaos: An Interdisciplinary Journal of Nonlinear Science {\bf 21}, 037107}~(2011).

\bibitem{crutchfield2009time}
James~P. Crutchfield, Christopher~J. Ellison, and John~R. Mahoney.
\newblock ``Time's barbed arrow: Irreversibility, crypticity, and stored information''.
\newblock \href{https://dx.doi.org/10.1103/PhysRevLett.103.094101}{Phys. Rev. Lett. {\bf 103}, 094101}~(2009).

\bibitem{nielsen2010quantum}
Michael~A. Nielsen and Isaac~L. Chuang.
\newblock ``Quantum {C}omputation and {Q}uantum {I}nformation''.
\newblock \href{https://dx.doi.org/10.1017/CBO9780511976667}{Cambridge University Press}. New York~(2000).

\bibitem{elliott2021memory}
Thomas~J. Elliott.
\newblock ``Memory compression and thermal efficiency of quantum implementations of nondeterministic hidden {M}arkov models''.
\newblock \href{https://dx.doi.org/10.1103/PhysRevA.103.052615}{Phys. Rev. A {\bf 103}, 052615}~(2021).

\bibitem{bosyk2012collision}
G.~M. Bosyk, M.~Portesi, and A.~Plastino.
\newblock ``Collision entropy and optimal uncertainty''.
\newblock \href{https://dx.doi.org/10.1103/PhysRevA.85.012108}{Phys. Rev. A {\bf 85}, 012108}~(2012).

\bibitem{csiszar1995generalized}
Imre Csisz{\'a}r.
\newblock ``Generalized cutoff rates and {R}{\'e}nyi's information measures''.
\newblock \href{https://dx.doi.org/10.1109/18.370121}{IEEE Transactions on information theory {\bf 41}, 26--34}~(1995).

\bibitem{bennett1995generalized}
Charles~H Bennett, Gilles Brassard, Claude Cr{\'e}peau, and Ueli~M Maurer.
\newblock ``Generalized privacy amplification''.
\newblock \href{https://dx.doi.org/10.1109/18.476316}{IEEE Transactions on Information theory {\bf 41}, 1915--1923}~(1995).

\bibitem{renner2008security}
Renato Renner.
\newblock ``Security of quantum key distribution''.
\newblock \href{https://dx.doi.org/10.1142/S0219749908003256}{International Journal of Quantum Information {\bf 6}, 1--127}~(2008).

\bibitem{csiszar1972class}
Imre Csisz{\'a}r.
\newblock ``A class of measures of informativity of observation channels''.
\newblock \href{https://dx.doi.org/10.1007/BF02018661}{Periodica Mathematica Hungarica {\bf 2}, 191--213}~(1972).

\bibitem{sibson1969information}
Robin Sibson.
\newblock ``Information radius''.
\newblock \href{https://dx.doi.org/10.1007/BF00537520}{Zeitschrift f{\"u}r Wahrscheinlichkeitstheorie und verwandte Gebiete {\bf 14}, 149--160}~(1969).

\bibitem{ho2015convexity}
Siu-Wai Ho and Sergio Verdú.
\newblock ``Convexity/concavity of {R\'e}nyi entropy and $\alpha$-mutual information''.
\newblock In 2015 IEEE International Symposium on Information Theory (ISIT).
\newblock Pages 745--749.
\newblock ~(2015).
\newblock  url:~\url{https://doi.org/10.1109/ISIT.2015.7282554}.

\bibitem{verdu2015alpha}
Sergio Verdú.
\newblock ``$\alpha$-mutual information''.
\newblock In 2015 Information Theory and Applications Workshop (ITA).
\newblock Pages 1--6.
\newblock ~(2015).
\newblock  url:~\url{https://doi.org/10.1109/ITA.2015.7308959}.

\bibitem{gallager1965simple}
R~Gallager.
\newblock ``A simple derivation of the coding theorem and some applications''.
\newblock \href{https://dx.doi.org/10.1109/TIT.1965.1053730}{IEEE Transactions on Information Theory {\bf 11}, 3--18}~(1965).

\bibitem{arimoto1973converse}
Suguru Arimoto.
\newblock ``On the converse to the coding theorem for discrete memoryless channels ({C}orresp.)''.
\newblock \href{https://dx.doi.org/10.1109/TIT.1973.1055007}{IEEE Transactions on Information Theory {\bf 19}, 357--359}~(1973).

\bibitem{polyanskiy2010arimoto}
Yury Polyanskiy and Sergio Verdú.
\newblock ``Arimoto channel coding converse and rényi divergence''.
\newblock In 2010 48th Annual Allerton Conference on Communication, Control, and Computing (Allerton).
\newblock Pages 1327--1333.
\newblock ~(2010).
\newblock  url:~\url{https://doi.org/10.1109/ALLERTON.2010.5707067}.

\bibitem{crutchfield2003regularities}
James~P Crutchfield and David~P Feldman.
\newblock ``Regularities unseen, randomness observed: Levels of entropy convergence''.
\newblock \href{https://dx.doi.org/10.1063/1.1530990}{Chaos: An Interdisciplinary Journal of Nonlinear Science {\bf 13}, 25--54}~(2003).

\bibitem{ellison2009prediction}
Christopher~J Ellison, John~R Mahoney, and James~P Crutchfield.
\newblock ``Prediction, retrodiction, and the amount of information stored in the present''.
\newblock \href{https://dx.doi.org/10.1007/s10955-009-9808-z}{Journal of Statistical Physics {\bf 136}, 1005--1034}~(2009).

\bibitem{renyi1961measures}
Alfr{\'e}d R{\'e}nyi et~al.
\newblock ``On measures of information and entropy''.
\newblock In Proceedings of the 4th Berkeley symposium on mathematics, statistics and probability, Volume 1: Contributions to the Theory of Statistics.
\newblock Pages 547--562.
\newblock University of California Press~(1961).
\newblock  url:~\url{https://projecteuclid.org/euclid.bsmsp/1200512181}.

\bibitem{daroczy1963gemeinsame}
Z~Dar{\'o}czy.
\newblock ``{\"U}ber die gemeinsame charakterisierung der zu den nicht vollst{\"a}ndigen verteilungen geh{\"o}rigen entropien von shannon und von r{\'e}nyi''.
\newblock \href{https://dx.doi.org/10.1007/BF00533413}{Zeitschrift f{\"u}r Wahrscheinlichkeitstheorie und verwandte Gebiete {\bf 1}, 381--388}~(1963).

\bibitem{brandenburger2025axiomatization}
Adam Brandenburger and Pierfrancesco~La Mura.
\newblock ``Axiomatization of {R}\'enyi entropy on quantum phase space''~(2025).
\newblock  \href{http://arxiv.org/abs/2410.15976}{arXiv:2410.15976}.

\bibitem{koukoulekidis2022constraints}
Nikolaos Koukoulekidis and David Jennings.
\newblock ``Constraints on magic state protocols from the statistical mechanics of {W}igner negativity''.
\newblock \href{https://dx.doi.org/10.1038/s41534-022-00551-1}{npj Quantum Information {\bf 8}, 42}~(2022).

\bibitem{wootters2007discrete}
William~K. Wootters and Daniel~M. Sussman.
\newblock ``Discrete phase space and minimum-uncertainty states''~(2007).
\newblock  \href{http://arxiv.org/abs/0704.1277}{arXiv:0704.1277}.

\bibitem{brukner2009information}
{\v{C}}aslav Brukner and Anton Zeilinger.
\newblock ``Information invariance and quantum probabilities''.
\newblock \href{https://dx.doi.org/10.1007/s10701-009-9316-7}{Foundations of Physics {\bf 39}, 677--689}~(2009).

\bibitem{brandenburger2022renyi}
Adam Brandenburger, Pierfrancesco La~Mura, and Stuart Zoble.
\newblock ``{R}{\'e}nyi entropy, signed probabilities, and the qubit''.
\newblock \href{https://dx.doi.org/10.3390/e24101412}{Entropy {\bf 24}, 1412}~(2022).

\bibitem{onggadinata2023qubits}
Kelvin Onggadinata, Pawe\l{} Kurzy\ifmmode~\acute{n}\else \'{n}\fi{}ski, and Dagomir Kaszlikowski.
\newblock ``Qubits from the classical collision entropy''.
\newblock \href{https://dx.doi.org/10.1103/PhysRevA.107.032214}{Phys. Rev. A {\bf 107}, 032214}~(2023).

\bibitem{rahimi2016sufficient}
Saleh Rahimi-Keshari, Timothy~C. Ralph, and Carlton~M. Caves.
\newblock ``Sufficient conditions for efficient classical simulation of quantum optics''.
\newblock \href{https://dx.doi.org/10.1103/PhysRevX.6.021039}{Phys. Rev. X {\bf 6}, 021039}~(2016).

\bibitem{koukoulekidis2022faster}
Nikolaos Koukoulekidis, Hyukjoon Kwon, Hyejung~H Jee, David Jennings, and MS~Kim.
\newblock ``Faster {B}orn probability estimation via gate merging and frame optimisation''.
\newblock \href{https://dx.doi.org/10.22331/q-2022-10-13-838}{Quantum {\bf 6}, 838}~(2022).

\bibitem{temme2017error}
Kristan Temme, Sergey Bravyi, and Jay~M. Gambetta.
\newblock ``Error mitigation for short-depth quantum circuits''.
\newblock \href{https://dx.doi.org/10.1103/PhysRevLett.119.180509}{Phys. Rev. Lett. {\bf 119}, 180509}~(2017).

\bibitem{takagi2021optimal}
Ryuji Takagi.
\newblock ``Optimal resource cost for error mitigation''.
\newblock \href{https://dx.doi.org/10.1103/PhysRevResearch.3.033178}{Phys. Rev. Res. {\bf 3}, 033178}~(2021).

\bibitem{takagi2022fundamental}
Ryuji Takagi, Suguru Endo, Shintaro Minagawa, and Mile Gu.
\newblock ``Fundamental limits of quantum error mitigation''.
\newblock \href{https://dx.doi.org/10.1038/s41534-022-00618-z}{npj Quantum Information {\bf 8}, 114}~(2022).

\bibitem{yuan2021universal}
Xiao Yuan, Yunchao Liu, Qi~Zhao, Bartosz Regula, Jayne Thompson, and Mile Gu.
\newblock ``Universal and operational benchmarking of quantum memories''.
\newblock \href{https://dx.doi.org/10.1038/s41534-021-00444-9}{npj Quantum Information {\bf 7}, 108}~(2021).

\bibitem{mitarai2021overhead}
Kosuke Mitarai and Keisuke Fujii.
\newblock ``Overhead for simulating a non-local channel with local channels by quasiprobability sampling''.
\newblock \href{https://dx.doi.org/10.22331/q-2021-01-28-388}{Quantum {\bf 5}, 388}~(2021).

\bibitem{veitch2014resource}
Victor Veitch, S~A~Hamed Mousavian, Daniel Gottesman, and Joseph Emerson.
\newblock ``The resource theory of stabilizer quantum computation''.
\newblock \href{https://dx.doi.org/10.1088/1367-2630/16/1/013009}{New Journal of Physics {\bf 16}, 013009}~(2014).

\bibitem{brunner2014bell}
Nicolas Brunner, Daniel Cavalcanti, Stefano Pironio, Valerio Scarani, and Stephanie Wehner.
\newblock ``Bell nonlocality''.
\newblock \href{https://dx.doi.org/10.1103/RevModPhys.86.419}{Rev. Mod. Phys. {\bf 86}, 419--478}~(2014).

\bibitem{clauser1969propesed}
John~F. Clauser, Michael~A. Horne, Abner Shimony, and Richard~A. Holt.
\newblock ``Proposed experiment to test local hidden-variable theories''.
\newblock \href{https://dx.doi.org/10.1103/PhysRevLett.23.880}{Phys. Rev. Lett. {\bf 23}, 880--884}~(1969).

\bibitem{popescu1994quantum}
Sandu Popescu and Daniel Rohrlich.
\newblock ``Quantum nonlocality as an axiom''.
\newblock \href{https://dx.doi.org/10.1007/BF02058098}{Foundations of Physics {\bf 24}, 379--385}~(1994).

\bibitem{ferrie2008frame}
Christopher Ferrie and Joseph Emerson.
\newblock ``Frame representations of quantum mechanics and the necessity of negativity in quasi-probability representations''.
\newblock \href{https://dx.doi.org/10.1088/1751-8113/41/35/352001}{Journal of Physics A: Mathematical and Theoretical {\bf 41}, 352001}~(2008).

\bibitem{ferrie2009framed}
Christopher Ferrie and Joseph Emerson.
\newblock ``Framed {H}ilbert space: {H}anging the quasi-probability pictures of quantum theory''.
\newblock \href{https://dx.doi.org/10.1088/1367-2630/11/6/063040}{New Journal of Physics {\bf 11}, 063040}~(2009).

\bibitem{ferrie2011quasi}
Christopher Ferrie.
\newblock ``Quasi-probability representations of quantum theory with applications to quantum information science''.
\newblock \href{https://dx.doi.org/10.1088/0034-4885/74/11/116001}{Reports on Progress in Physics {\bf 74}, 116001}~(2011).

\bibitem{kofler2007classical}
Johannes Kofler and {\v{C}}aslav Brukner.
\newblock ``Classical world arising out of quantum physics under the restriction of coarse-grained measurements''.
\newblock \href{https://dx.doi.org/10.1103/PhysRevLett.99.180403}{Phys. Rev. Lett. {\bf 99}, 180403}~(2007).

\bibitem{oas2014exploring}
G~Oas, J~Acacio de~Barros, and C~Carvalhaes.
\newblock ``Exploring non-signalling polytopes with negative probability''.
\newblock \href{https://dx.doi.org/10.1088/0031-8949/2014/T163/014034}{Physica Scripta {\bf 2014}, 014034}~(2014).

\bibitem{marzen2015informational}
Sarah~E. Marzen and James~P. Crutchfield.
\newblock ``Informational and causal architecture of discrete-time renewal processes''.
\newblock \href{https://dx.doi.org/10.3390/e17074891}{Entropy {\bf 17}, 4891--4917}~(2015).

\bibitem{mahoney2009information}
John~R Mahoney, Christopher~J Ellison, and James~P Crutchfield.
\newblock ``Information accessibility and cryptic processes''.
\newblock \href{https://dx.doi.org/10.1088/1751-8113/42/36/362002}{Journal of Physics A: Mathematical and Theoretical {\bf 42}, 362002}~(2009).

\bibitem{perron1907theorie}
Oskar Perron.
\newblock ``Zur theorie der matrices''.
\newblock \href{https://dx.doi.org/10.1007/BF01449896}{Mathematische Annalen {\bf 64}, 248--263}~(1907).

\bibitem{frobenius1912matrizen}
Georg Frobenius.
\newblock ``{\"U}ber {M}atrizen aus nicht negativen {E}lementen''.
\newblock K{\"o}nigliche Akademie der Wissenschaften Sitzungsber, K{\"o}n. ~(1912).
\newblock  url:~\url{https://upload.wikimedia.org/wikipedia/commons/4/44/Ueber_Matrizen_aus_nicht_negativen_Elementen.pdf}.

\bibitem{tarazaga2001perron}
Pablo Tarazaga, Marcos Raydan, and Ana Hurman.
\newblock ``Perron–{F}robenius theorem for matrices with some negative entries''.
\newblock \href{https://dx.doi.org/https://doi.org/10.1016/S0024-3795(00)00327-X}{Linear Algebra and its Applications {\bf 328}, 57--68}~(2001).

\bibitem{johnson2004matrices}
Charles~R Johnson and Pablo Tarazaga.
\newblock ``On matrices with {P}erron-{F}robenius properties and some negative entries''.
\newblock \href{https://dx.doi.org/10.1007/s11117-003-3881-3}{Positivity {\bf 8}, 327--338}~(2004).

\bibitem{noutsos2006perron}
Dimitrios Noutsos.
\newblock ``On {P}erron–{F}robenius property of matrices having some negative entries''.
\newblock \href{https://dx.doi.org/https://doi.org/10.1016/j.laa.2005.06.037}{Linear Algebra and its Applications {\bf 412}, 132--153}~(2006).

\bibitem{curgus2015somewhat}
Branko {\'C}urgus and Robert~I Jewett.
\newblock ``Somewhat stochastic matrices''.
\newblock \href{https://dx.doi.org/10.4169/amer.math.monthly.122.01.36}{The American Mathematical Monthly {\bf 122}, 36--42}~(2015).

\bibitem{chruscinski2015pseudo}
D~Chruściński, V~I Man’ko, G~Marmo, and F~Ventriglia.
\newblock ``On pseudo-stochastic matrices and pseudo-positive maps''.
\newblock \href{https://dx.doi.org/10.1088/0031-8949/90/11/115202}{Physica Scripta {\bf 90}, 115202}~(2015).

\bibitem{wootters1987wigner}
William~K Wootters.
\newblock ``A {W}igner-function formulation of finite-state quantum mechanics''.
\newblock \href{https://dx.doi.org/https://doi.org/10.1016/0003-4916(87)90176-X}{Annals of Physics {\bf 176}, 1--21}~(1987).

\bibitem{pawlowski2009information}
Marcin Paw{\l}owski, Tomasz Paterek, Dagomir Kaszlikowski, Valerio Scarani, Andreas Winter, and Marek {\.Z}ukowski.
\newblock ``Information causality as a physical principle''.
\newblock \href{https://dx.doi.org/10.1038/nature08400}{Nature {\bf 461}, 1101--1104}~(2009).

\bibitem{cabello2018optimal}
Ad\'an Cabello, Mile Gu, Otfried G\"uhne, and Zhen-Peng Xu.
\newblock ``Optimal classical simulation of state-independent quantum contextuality''.
\newblock \href{https://dx.doi.org/10.1103/PhysRevLett.120.130401}{Phys. Rev. Lett. {\bf 120}, 130401}~(2018).

\bibitem{aghamohammadi2017ambiguity}
Cina Aghamohammadi, John~R Mahoney, and James~P Crutchfield.
\newblock ``The ambiguity of simplicity in quantum and classical simulation''.
\newblock \href{https://dx.doi.org/https://doi.org/10.1016/j.physleta.2016.12.036}{Physics Letters A {\bf 381}, 1223--1227}~(2017).

\bibitem{breuer1995stochastic}
Heinz-Peter Breuer and Francesco Petruccione.
\newblock ``Stochastic dynamics of open quantum systems: Derivation of the differential chapman-kolmogorov equation''.
\newblock \href{https://dx.doi.org/10.1103/PhysRevE.51.4041}{Phys. Rev. E {\bf 51}, 4041--4054}~(1995).

\bibitem{milz2021quantum}
Simon Milz and Kavan Modi.
\newblock ``Quantum stochastic processes and quantum non-markovian phenomena''.
\newblock \href{https://dx.doi.org/10.1103/PRXQuantum.2.030201}{PRX Quantum {\bf 2}, 030201}~(2021).

\bibitem{garner2017thermodynamics}
Andrew J.~P. Garner, Jayne Thompson, Vlatko Vedral, and Mile Gu.
\newblock ``Thermodynamics of complexity and pattern manipulation''.
\newblock \href{https://dx.doi.org/10.1103/PhysRevE.95.042140}{Phys. Rev. E {\bf 95}, 042140}~(2017).

\bibitem{chiribella2009theoretical}
Giulio Chiribella, Giacomo~Mauro D'Ariano, and Paolo Perinotti.
\newblock ``Theoretical framework for quantum networks''.
\newblock \href{https://dx.doi.org/10.1103/PhysRevA.80.022339}{Phys. Rev. A {\bf 80}, 022339}~(2009).

\bibitem{oreshkov2012quantum}
Ognyan Oreshkov, Fabio Costa, and {\v{C}}aslav Brukner.
\newblock ``Quantum correlations with no causal order''.
\newblock \href{https://dx.doi.org/https://doi.org/10.1038/ncomms2076}{Nature communications {\bf 3}, 1092}~(2012).

\end{thebibliography}

\ULforem

\onecolumngrid

\begin{appendix}
\section{Proofs for Section \ref{sec: alternative measures}}\label{sec: appendix proof new measures}

We first show that $I_{\frac{1}{2}}[\mathcal{S};\Future]=\mathbf{E}_{\frac{1}{2}}$:
\begin{eqnarray}
I_{\frac{1}{2}}[\mathcal{S};\Future] &=& -\log\sum_{\future}\left(\sum_{k}P(\sigma_k)\sqrt{P(\future|\sigma_k)}\right)^2 \nonumber \\
&=& -\log\sum_{\future}\left(\sum_{k}\sum_{\past\in\sigma_k}P(\past)\sqrt{P(\future|\sigma_k)}\right)^2 \nonumber \\
&=& -\log\sum_{\future}\left(\sum_{\past\in\sigma_1}P(\past)\sqrt{P(\future|\sigma_1)} + \dots + \sum_{\past\in\sigma_{|\bm{\mathcal{S}}|}}P(\past)\sqrt{P(\future|\sigma_{|\bm{\mathcal{S}}|})}\right)^2 \nonumber \\
&=& -\log\sum_{\future}\left(\sum_{\past\in\sigma_1}P(\past)\sqrt{P(\future|\past)} + \dots + \sum_{\past\in\sigma_{|\bm{\mathcal{S}}|}}P(\past)\sqrt{P(\future|\past)}\right)^2 \nonumber \\
&=& -\log\sum_{\future}\left(\sum_{\past}P(\past)\sqrt{P(\future|\past)} \right)^2 \nonumber
\end{eqnarray}
Here, we have use the fact that $P(\sigma_k) = \sum_{\past \in \sigma_k}P(\past)$ and $P(\future|\sigma_k)=P(\future|\past)\, \forall \,\past\in\sigma_k$. 

Next, we show the proof of the inequality $C_\mu^{(2)} \geq C_q^{(2)} \geq \mathbf{E}_{\frac{1}{2}}$. As mentioned in the main text, Ref. \cite{ho2015convexity} already proved that $C_\mu^{(2)} \geq \mathbf{E}_{\frac{1}{2}}$. Furthermore, it is easy to see that 
\begin{eqnarray}
C_q^{(2)} \equiv S_2[\rho] &=& -\log\tr{\sum_{kl} \pi_k\pi_l\ket{\sigma_k}\braket{\sigma_k|\sigma_l}\bra{\sigma_l}} \nonumber \\
&=& -\log\sum_{kl}\pi_k\pi_l |\braket{\sigma_k|\sigma_l}|^2 \nonumber \\
&\leq & -\log\sum_{kl} \pi_k\pi_l \delta_{kl} \nonumber \\
&=& -\log\sum_{k} \pi_k^2\, ,
\end{eqnarray}
which tells us that $C_\mu^{(2)} \geq C_q^{(2)}$. The inequality in the third line is obtained since we have $\sum_{k}\pi_k^2 \leq \sum_{k}\pi_k^2 + \sum_{k\neq l} \pi_k\pi_l|\braket{\sigma_k|\sigma_l}|^2$ with equality iff $\{\ket{\sigma_k}\}$ are mutually orthogonal, which it is not the case generally. Lastly, to show that $C_q^{(2)} \geq \mathbf{E}_{\frac{1}{2}}$, we use the relation from \cite{binder2018practical} that the overlap between the q-machine's states is equal to 
\begin{equation}
     \braket{\sigma_j|\sigma_k} = \sum_{\future}\sqrt{P(\future|\sigma_j)P(\future|\sigma_k)}\, .
\end{equation}
This implies that
\begin{eqnarray}
I_{\frac{1}{2}}[\mathcal{S}_q;\Future] &=& -\log\sum_{\future}\left(\sum_k P(\sigma_k)\sqrt{P(\future|\sigma_k)}\right)^2 \nonumber \\
&=& -\log\sum_{j,k} P(\sigma_j)P(\sigma_k)\sum_{\future}\sqrt{P(\future|\sigma_j)P(\future|\sigma_k)} \nonumber \\
&=& -\log\sum_{j,k} P(\sigma_j)P(\sigma_k) |\braket{\sigma_j|\sigma_k}| \nonumber \\
&\leq & -\log\sum_{j,k} P(\sigma_j)P(\sigma_k) |\braket{\sigma_j|\sigma_k}|^2 \nonumber \\
&=& -\log \tr{\left(\sum_k \pi_k \ket{\sigma_k}\bra{\sigma_k}\right)^2} = S_2[\rho]\, .
\end{eqnarray}
Therefore, we have completed the proof for $C_\mu^{(2)} \geq C_q^{(2)} \geq \mathbf{E}_{\frac{1}{2}}$. Note that this also works for different protocol of constructing q-machine. For instance, Liu {\it et al.}'s phase-optimized q-simulator \cite{liu2019optimal} and Elliott {\it et al.}'s extreme compression \cite{elliott2020extreme} has slightly different form of $U$ than the one in Eq. \eqref{eq: q machine unitary}, but it can be shown easily that it will yield the same results as above.

The final proof we will show in this section is the cases of stochastic process where $\mathbf{E}_{\frac{1}{2}} = I_{\frac{1}{2}}[\mathcal{S}^+; \mathcal{S}^-]$. The first is the Perturbed Coin Process with the $\epsilon$-machine representation shown in Fig. \ref{fig: perturbed coin epsilon machine}. Here, the forward and reverse causal states are equal, i.e. $\mathcal{S}^+=\mathcal{S}^-$, because $P(\Future|\Past) = P(\Past^-|\Future^-)$ where $\Past^- = X_{-1}X_{-2}\dots$ and $\Future^- = X_0X_1X_2\dots$. So $\mathcal{S}^- = \{\sigma_0^-,\sigma_1^-\}$ where $\sigma_0^- = \{\future : \textup{$\future$ starts with $0$}\}$ and $\sigma_1^- = \{\future : \textup{$\future$ starts with $1$}\}$. Hence, this implies that
\begin{eqnarray}
I_\frac{1}{2}[\mathcal{S}^+; \mathcal{S}^-] &=& I_{\frac{1}{2}}[\Past; \mathcal{S}^-] \nonumber \\
&=& -\log\sum_{\sigma^-} \bigg( \sum_{\past} P_{\Past}(\past) \sqrt{P_{\mathcal{S}^-|\Past}(\sigma^-|\past)} \bigg)^2 \nonumber \\
&=& -\log \sum_{\past,\past'} P_{\Past}(\past)P_{\Past}(\past') \sum_{\sigma^-} \sqrt{\sum_{\future,\future'\in \sigma^-} P_{\Future|\Past}(\future|\past)P_{\Future|\Past}(\future'|\past')} \nonumber \\
&=& -\log \sum_{\past,\past'} P_{\Past}(\past)P_{\Past}(\past') \sum_{\past} \sqrt{P_{\Future|\Past}(\future|\past)P_{\Future|\Past}(\future|\past')} \nonumber \\
&=& I_{\frac{1}{2}}[\Past;\Future]\, .,
\end{eqnarray}
where we have used the notation $P_{X}(x)=P(X=x)$ for clarity. In the above, line 4 follows from
\begin{equation}
P(\sigma^-|\past) = \sum_{\future\in \sigma^-} P(\future|\past)
= \sum_{x_{n+1}x_{n+3}\dots} P(x_n|\past) P(x_{n+1}x_{n+3}\dots|\past x_n) = P(x_n|\past)\, ,
\end{equation}
where $x_n=0$ if $\sigma^-=\sigma_0^-$ and $x_n=1$ if $\sigma^-=\sigma_1^-$. Using the same basic idea, this can also be proven for the Golden Mean Process (order-1 Markov) \cite{mahoney2009information}.

The second process with $\mathbf{E}_{\frac{1}{2}} = I_{\frac{1}{2}}[\mathcal{S}^+; \mathcal{S}^-]$ is the SNS Process with its (forward-time) HMM representation given in Fig. \ref{fig: sns hmm representation}. In this case, the forward and backward causal states are $\bm{\mathcal{S}}^+ = \{\sigma_n^+\}$, $\bm{\mathcal{S}}^- = \{\sigma_n^-\}$, where $\sigma_n^+$ contains past $\past$ that ends with $10^n$ and $\sigma_n^-$ contains past that starts with $0^n1$. Therefore,
\begin{align*}
& I_\frac{1}{2}[\mathcal{S}^+;\mathcal{S}^-] = I_\frac{1}{2}[\Past;\mathcal{S}^-] \\
&= -\log\sum_{\sigma^-} \bigg( \sum_{\past} P_\Past(\past) \sqrt{P_{\mathcal{S}^-|\Past}(\sigma^-|\past)} \bigg)^2 \\
&= -\log \sum_{\past,\past'} P_\Past(\past)P_\Past(\past') \sum_{\sigma^-} \sqrt{\sum_{\future,\future'\in \sigma^-} P_{\Future|\Past}(\future|\past)P_{\Future|\Past}(\future'|\past')} \\
&= -\log \sum_{\past,\past'} P_\Past(\past)P_\Past(\past') \sum_{n=0}^\infty \sqrt{\sum_{\future,\future'\in \sigma_n^-} P_{\Future|\Past}(\future|\past)P_{\Future|\Past}(\future'|\past')} \\
&= -\log \sum_{\past,\past'} P_\Past(\past)P_\Past(\past')\sum_{n=0}^\infty \sqrt{P_{\Future^{n+1}|\Past}(0^n1|\past)P_{\Future^{n+1}|\Past}(0^n1|\past') \sum_{\future,\future'} P_{\Future|\Past}(\future|\past0^n1)P_{\Future|\Past}(\future'|\past'0^n1) } \\
&= -\log \sum_{\past,\past'} P_\Past(\past)P_\Past(\past') \sum_{n=0}^\infty \sqrt{P_{\Future^{n+1}|\Past}(0^n1|\past)P_{\Future^{n+1}|\Past}(0^n1|\past')} \\
&= -\log \sum_{\past,\past'} P_\Past(\past)P_\Past(\past') \sum_{n=0}^\infty \sum_{x_{n+1}x_{n+2}\dots} P_{\Future|\Past}(x_{n+1}x_{n+2}\dots|1)\sqrt{P_{\Future^{n+1}|\Past}(0^n1|\past)P_{\Future^{n+1}|\Past}(0^n1|\past')} \\
&= -\log \sum_{\past,\past'} P_\Past(\past)P_\Past(\past')  \\
&\qquad\times\sum_{n=0}^\infty \sum_{x_{n+1}x_{n+2}\dots} \sqrt{P_{\Future^{n+1}|\Past}(0^n1|\past)P_{\Future|\Past}(x_{n+1}x_{n+2}\dots|1)P_{\Future^{n+1}|\Past}(0^n1|\past')P_{\Future|\Past}(x_{n+1}x_{n+2}\dots|1)} \\
&= -\log \sum_{\past,\past'} P_\Past(\past)P_\Past(\past') \sum_{\past} \sqrt{P_{\Future|\Past}(\future|\past)P_{\Future|\Past}(\future|\past')} \\
&= I_\frac{1}{2}[\Past:\Future] \, ,
\end{align*}
where equality 5-9 above follows from
\begin{align*}
P(\sigma_n^-|\past) &= \sum_{\future\in \sigma_n^-} P(\future|\past) \\
&= \sum_{x_{n+1}x_{n+3}\dots} P(0^n1x_{n+1}x_{n+3}\dots|\past) \\
&= \sum_{x_{n+1}x_{n+3}\dots} P(0^n1|\past) P(x_{n+1}x_{n+3}\dots|\past0^n1) \\
&= P(0^n1|\past) \sum_{x_{n+1}x_{n+3}\dots} P(x_{n+1}x_{n+3}\dots|1) = P(0^n1|\past)
\end{align*}
because $P(x_{n+1}x_{n+3}\dots|\past1) = P(x_{n+1}x_{n+3}\dots|1)$ for all $\past$.
    
\section{Proof of properties in Section \ref{sec: n machine}}\label{sec: appendix proof n-machine}

To prove Eq. \eqref{eq: n-machine construct 3}, notice that:
\begin{equation}
\sum_{l_k}\tilde{\pi}_{k,l_k} = \sum_{l_k}\sum_{j,l_j} \tilde{T}_{j,l_j;k,l_k}\tilde{\pi}_{j,l_j} = \sum_{j}T_{j;k}\sum_{l_j}\tilde{\pi}_{j,l_j}
\end{equation}
where we have used Eq. \eqref{eq: n-machine construct 1} written as $\sum_{l_k,x}\tilde{T}^{(x)}_{j,l_j;k,l_k} = \sum_{l_k}\tilde{T}_{j,l_j;k,l_k} = T_{j;k}\, \forall \, k,j,l_j$. Since this is equivalent to Eq. \eqref{eq: stationary distribution eigenproblem}, it implies that $\sum_{l_k}\tilde{\pi}_{k,l_k} = \pi_k$.

To prove Eq. \eqref{eq: n-machine construct 2 extended}, we first show that
\begin{eqnarray}
    P(\future^L|\tilde{\sigma}_{k,l_k}) &=& \sum_{k_1, l_{k_1}}\sum_{k_2, l_{k_2}} \dots \sum_{k_{L-1},l_{k_{L-1}}} P(x_L|\tilde{\sigma}_{k_{L-1},l_{k_{L-1}}})P(\tilde{\sigma}_{k_{L-1},l_{k_{L-1}}},x_{L-1}|\tilde{\sigma}_{k_{L-2},l_{k_{L-2}}}) \dots P(\tilde{\sigma}_{k_1,l_{k_1}},x_1|\tilde{\sigma}_{k,l_k}) \nonumber \\
    &=& \sum_{k_1, l_{k_1}}\sum_{k_2, l_{k_2}} \dots \sum_{k_{L-1}} P(x_L|\sigma_{k_{L-1}})\sum_{l_{k_{L-1}}}P(\tilde{\sigma}_{k_{L-1},l_{k_{L-1}}},x_{L-1}|\tilde{\sigma}_{k_{L-2},l_{k_{L-2}}}) \dots P(\tilde{\sigma}_{k_1,l_{k_1}},x_1|\tilde{\sigma}_{k,l_k}) \nonumber \\
    &=& \sum_{k_1, l_{k_1}}\sum_{k_2, l_{k_2}} \dots \sum_{k_{L-1}} P(x_L|\sigma_{k_{L-1}})P(\sigma_{k_{L-1}},x_{L-1}|\tilde{\sigma}_{k_{L-2}}) \dots P(\tilde{\sigma}_{k_1,l_{k_1}},x_1|\tilde{\sigma}_{k,l_k}) \nonumber \\
    & & \vdots \nonumber \\
    &=& \sum_{k_1,k_2, \dots ,k_{L-1}} P(x_L|\sigma_{k_{L-1}})P(\sigma_{k_{L-1}},x_{L-1}|\tilde{\sigma}_{k_{L-2}})\dots P(\sigma_{k_1},x_1|\sigma_{k}) \nonumber \\
    &=& P(\future^L|\sigma_k)\, ,
\end{eqnarray}
where in the second line we have used the property in Eq. \eqref{eq: n-machine construct 2}. In going to the third line, we have used the construction in Eq. \eqref{eq: n-machine construct 1}. Subsequent simplification follows the same idea. Since this works for any $L$, we can conclude that $P(\future|\tilde{\sigma}_{k,l_k}) = P(\future|\sigma_k)$ for all $\future,k,l_k$. 

From here it is straightforward to show that the n-machine constructed from the protocol above will produce the same stochastic process as its previous $\epsilon$-machine. We say that an $\epsilon$-machine $(\mathcal{S},\mathcal{T},\pi)$ and an n-machine $(\tilde{\mathcal{S}},\tilde{\mathcal{T}},\tilde{\pi})$ is equivalent if they produces identical word realization $w= \future^L =x_1x_2\dots x_L$:
\begin{equation}
    \underbrace{\pi T^{(x_1)}T^{(x_2)}\dots T^{(x_L)}\mathbf{1}}_{\equiv P_{\epsilon}(w)} = \underbrace{\tilde{\pi}\tilde{T}^{(x_1)}\tilde{T}^{(x_2)}\dots \tilde{T}^{(x_L)}\mathbf{1}}_{\equiv P_n(w)}\quad \forall \, w\in\mathcal{A}^L\, ,
\end{equation}
where $\mathbf{1} = [1,1,\dots,1]^T$, and we denote $P_\epsilon(w)$ as the word realization generated by $\epsilon$-machine. Similarly, $P_n(w)$ is the realization by n-machine. Their equivalence can be seen clearly from 
\begin{eqnarray}
P_n(w) & \equiv & \sum_{k,l_k}P(\future^L|\tilde{\sigma}_{k,l_k})P(\tilde{\sigma}_{k,l_k}) \nonumber \\
&=& \sum_{k}P(\future^L|\sigma_k)\sum_{l_k}P(\tilde{\sigma}_{k,l_k}) \nonumber \\
&=& \sum_{k}P(\future^L|\sigma_k)P(\sigma_k) = P_{\epsilon}(w)\, .
\end{eqnarray}
Here, we have use the fact that Eq. \eqref{eq: n-machine construct 3} holds.

Proof that $I_{\frac{1}{2}}[\tilde{\mathcal{S}}; \Future] = I_{\frac{1}{2}}[\mathcal{S};\Future]$:
\begin{eqnarray}
I_{\frac{1}{2}}[\tilde{\mathcal{S}};\Future] &=& -\log\sum_{\future}\left(\sum_{k,l_k}P(\tilde{\sigma}_{k,l_k})\sqrt{P(\future|\tilde{\sigma}_{k,l_k})})\right)^2 \nonumber \\
&=& -\log\sum_{\future}\left(\sum_{k,l_k}P(\tilde{\sigma}_{k,l_k})\sqrt{P(\future|\sigma_{k})})\right)^2 \nonumber \\
&=& -\log\sum_{\future}\left(\sum_{k}P(\sigma_{k})\sqrt{P(\future|\sigma_{k})})\right)^2 = I_{\frac{1}{2}}[\mathcal{S};\Future]
\end{eqnarray}
which then also implies that $I_{\frac{1}{2}}[\tilde{\mathcal{S}};\Future] = \mathbf{E}_{\frac{1}{2}}$. Thus, n-machines also satisfy data processing inequality. Note that this can also be shown using the standard definition of excess entropy, i.e., $I[\tilde{\mathcal{S}};\Future] = I[\mathcal{S};\Future] = \mathbf{E}$.

\section{Calculation details for SNS Process}\label{sec: appendix examples details}

Here, we will provide the closed-form expression for $\mathbf{E}_\frac{1}{2}$ for SNS Process. Note first that we have shown the equivalence between $\mathbf{E}_\frac{1}{2} = I_{\frac{1}{2}}[\Past;\Future]$ and $I_{\frac{1}{2}}[\mathcal{S}^+; \mathcal{S}^-]$ for the SNS Process in Appendix \ref{sec: appendix proof new measures}. Moreover, from \cite{marzen2015informational}, it has been found that 
\begin{equation}
    P(\sigma_n^+, \sigma_m^-) = \mu \phi(m+n) \quad \text{and} \quad P(\sigma_n^+|\sigma^-_m) = \frac{\phi(m+n)}{\Phi(m)}\, .
\end{equation}
Therefore, we have
\begin{eqnarray}
I_{\frac{1}{2}}[\mathcal{S}^+; \mathcal{S}^-] &=& -\log\sum_{m=0}^\infty \left(\sum_{n=0}^\infty P(\sigma^+_n)\sqrt{P(\sigma^-_m|\sigma^+_n)} \right)^2 \nonumber \\
&=& -\log\sum_{m=0}^\infty \left(\sum_{n=0}^\infty \sqrt{P(\sigma^+_n,\sigma^-_m)P(\sigma^+_n)} \right)^2 \nonumber \\
&=& -\log\sum_{m=0}^\infty \left(\sum_{n=0}^\infty \mu \sqrt{ \phi(m+n)\Phi(n)} \right)^2
\end{eqnarray}
The term inside the $\log$ is exactly the quantity $M$ in Eq. \eqref{eq: quantity inside log excess entropy sns process}.

\section{GPT HMM through linear map}\label{sec: appedix gpt hmm linear map}

In this section, we provide more details regarding the general method of finding other equivalent HMM (even in GPT) through a linear invertible map $Z$. Given a stochastic process $P(\PastFuture)$ with a generator $g_1\equiv(\mathcal{A}, \bm{\mathcal{S}}, \{T^{(x)}\}, \pi)$, one can find another generator $g_2 \equiv (\mathcal{A}, \bm{\mathcal{S}}', \{T'^{(x)}\}, \pi')$ such that transformations are given by 
\begin{eqnarray}
 \pi' &=& \pi Z^{-1}\, , \\
T'^{(x)} &=&  ZT^{(x)}Z^{-1}\, ,\\
\bm{1} &=& Z\bm{1}\, .
\end{eqnarray}
It is clear that $g_2$ will generate the same stochastic process as $g_1$ from the relations 
\begin{eqnarray}
P_{g_{2}}(w) \equiv \pi' T'^{(x_1)}T'^{(x_2)}\dots T'^{(x_L)}\bm{1} &=& \pi Z^{-1}Z T^{(x_1)}Z^{-1}ZT^{(x_2)}Z^{-1}Z\dots Z^{-1}ZT^{(x_L)}Z^{-1}Z\bm{1} \nonumber \\
&=& \pi T^{(x_1)}T^{(x_2)}\dots T^{(x_L)}\bm{1} \equiv P_{g_1}(w) 
\end{eqnarray}
for all $w\in \mathcal{A}^L$ and any length $L$. Here, we have only assume that $Z$ is an invertible matrix, and following the above constraint we know that $Z$ is a $|\bm{\mathcal{S}}'|\times|\bm{\mathcal{S}}|$ real matrix and is (row) quasi-stochastic $\sum_{k}Z_{jk}=1\, \forall\, j$.

Let us use this approach to find the optimal generator model for Perturbed Coin (i.e. RJMC model in Fig. \ref{fig: perturbed coin rjmc model}) given that we know that $\epsilon$-machine. Assuming that $|\bm{\mathcal{S}}'|$ is equal to $|\bm{\mathcal{S}}|=2$, the general map $Z$ takes the form
\begin{equation}
    Z = \begin{bmatrix}
        a & 1-a \\ b & 1-b
    \end{bmatrix}
\end{equation}
such that $a\neq b$ so that $Z$ is invertible. To optimize it over the classical generator model, we require that $T'^{(x)}$ to have strictly non-negative elements for all $x\in\mathcal{A}$. Thus, this imposes a constraint on the domain of $a,b$:
\begin{eqnarray}
0 < p < \frac{1}{2}&:&\begin{cases}\frac{p}{-1+2p} \leq a \leq 0 \text{ and } 1 \leq b \leq \frac{-1+p}{-1+2p}\, ,\\ 1 \leq a \leq \frac{-1+p}{-1+2p} \text{ and } \frac{p}{-1+2p} \leq b \leq 0 \, ,\end{cases} \\
\frac{1}{2} < p < 1 &:& \begin{cases}\frac{-1+p}{-1+2p} \leq a \leq 0 \text{ and } 1 \leq b \leq \frac{p}{-1+2p}\, , \\ 1\leq a \leq \frac{p}{-1+2p} \text{ and } \frac{-1+p}{-1+2p}\leq b \leq 0\, .\end{cases}
\end{eqnarray}
It can be seen easily that when one picks $a=\frac{p}{-1+2p},b=1$ for $p\in(0,\frac{1}{2})$ and $a=\frac{-1+p}{-1+2p},b=1$ for $p\in(\frac{1}{2}, 1)$, one recovers exactly the RJMC model. 

Now, if we do not restrict ourselves to classical model, then we can still use this approach to find non-classical model with negative transition probabilities. As a matter of fact we can find a non-classical 2-state model with negative transition probabilities and positive stationary distribution. This can be achieved by $Z$ with the solution set
\begin{equation}
\begin{cases}
    a<\frac{1}{2} & \text{ and } b\geq \frac{1}{2}, \text{ or} \\
    a = \frac{1}{2} & \text{ and } b\neq \frac{1}{2}, \text{ or} \\
    a > \frac{1}{2} & \text{ and } b\leq \frac{1}{2}. \\
\end{cases}
\end{equation}
If for instance we take $a>\frac{1}{2}$ and $b=0$, this gives us
\begin{eqnarray}
T'^{(0)} &=& \begin{bmatrix}
    1 + (-2+\frac{1}{a})p & -1+a+3p-\frac{p}{a}-2ap \\ \frac{p}{a} & \frac{(-1+a)p}{a}
\end{bmatrix}\, , \\ 
T'^{(1)} &=& \begin{bmatrix}
    0 & 1-p+a(-1+2p) \\ 0 & 1-p
\end{bmatrix}\, ,\\ 
\pi' &=& \begin{bmatrix}
    \frac{1}{2a} & \frac{-1+2a}{2a}
\end{bmatrix}\, ,
\end{eqnarray}
where $T'{(x)}$ now contains quasiprobabilities and $\pi'$ is always positive when $a>\frac{1}{2}$. Moreover, as $a\to \infty$ we have $H[\pi'] \to 0$. Although this seems to be superior in terms of memory compression than the model obtained through the n-machine construction, we found that it is not meaningful to calculate the predictive information $I_{\frac{1}{2}}[\mathcal{S}';\Future]$ as it yields complex values. This is in contrast to the n-machines, where $I_{\frac{1}{2}}[\tilde{\mathcal{S}};\Future]$ (and even $I[\tilde{\mathcal{S}};\Future]$) can be calculated properly, and thus data processing inequality can be shown to be obeyed. 

Using this approach, it may provide us a way to find another HMM in the GPT region that is more unrestricted than the n-machine construction allows. As such, it is possible to find a model with seemingly extreme compression but fundamental properties such as data processing inequality cannot be shown to be satisfied at this point.

\section{Quasiprobability representation of HQMM}\label{sec: appendix qpr}

In this section, we provide a brief introduction to quasiprobability representation of quantum mechanics, and use an example to highlight a few comparison with the n-machine model.

Quasiprobability representation (or frame representation) is an alternative mathematical framework to describe elements of quantum mechanics in terms of objects in (quasi-)probability theory instead of the textbook Hilbert space formalism. The main tool is {\it frame} $\{F_\lambda\}_{\lambda\in \Lambda}$ and {\it dual frame} $\{G_\lambda\}_{\lambda\in\Lambda}$, which are overcomplete bases for the Hermitian space $\mathbb{H}(\mathcal{H}) = \{A\in \mathbb{C}^{d\times d}\,|\, A^\dagger = A\}$ such that they satisfy $A = \sum_{\lambda \in \Lambda} \tr{AF_\lambda}G_\lambda\, \forall A\in \mathbb{H}(\mathcal{H})$. Using this, one can represent a quantum state $\rho$, quantum channel $\Phi$, and elements of POVM $E_k$ as quasiprobability distribution on $\Lambda$ by the following map:
\begin{eqnarray}
\mu^\rho(\lambda) &=& \tr{F_\lambda \rho} \, ,\\
\tau^\Phi(\lambda|\lambda') &=& \tr{F_\lambda \Phi[G_{\lambda'}]} \, ,\\
\nu^{E_k}(\lambda) &=& \tr{G_\lambda E_k}\, .
\end{eqnarray}
More properties are imposed on the frame and dual frame's operators which we refer to \cite{ferrie2011quasi} for a more complete review. 

\begin{figure}[!htb]
\centering
\resizebox{0.8\textwidth}{!}{
    \begin{tikzpicture}[->, >=stealth', auto, semithick, node distance=3.5cm]
    \tikzstyle{every state}=[fill=white,draw=black,thick,text=black,scale=1]
    \node[draw=none, fill=none] (x) {};
    \node[state]    (A)[above of=x]         {$\tilde{\sigma}_{0,0}$};
    \node[state]    (B)[right of=x]   		{$\tilde{\sigma}_{0,1}$};
    \node[state]    (C)[below of=x]         {$\tilde{\sigma}_{1,0}$};
    \node[state]    (D)[left of=x]          {$\tilde{\sigma}_{1,1}$};
    \path
    (A) 
        edge[loop above, above] 	node{$0|\frac{1-\chi_-}{2}$}    (A)
        edge[right] 	node[pos=0.35]{$0|\frac{1-\chi_+}{2}$}    (B)
        edge[bend left, below] node[right]{$0|\frac{\chi_+}{2}$}  (C)
        edge[bend right, left] node{$0|\frac{\chi_-}{2}$}  (D)
    (B)	
        edge[bend right, above, red] 	node[right,pos=0.75]{$0|\frac{1-\chi_-}{2}$}    (A)
        edge[loop right, right, red] 	node{$0|\frac{1-\chi_+}{2}$}    (B)
        edge[below, red] node[right]{$0|\frac{\chi_+}{2}$}  (C)
        edge[bend left, left, red] node[below]{$0|\frac{\chi_-}{2}$}  (D)
    (C) 
        edge[bend left, above, blue] 	node[left]{$1|\frac{\chi_+}{2}$}    (A)
        edge[bend right, right, blue] 	node{$1|\frac{\chi_-}{2}$}    (B)
        edge[pos=0.4][loop below, below, blue] node{$1|\frac{1-\chi_-}{2}$}  (C)
        edge[left, blue] node[pos=0.45]{$1|\frac{1-\chi_+}{2}$}  (D)
    (D) 
        edge[above, OliveGreen] 	node[left]{$1|\frac{\chi_+}{2}$}    (A)
        edge[bend left, right, OliveGreen] 	node[above]{$1|\frac{\chi_-}{2}$}    (B)
        edge[bend right, below, OliveGreen] node[left,pos=0.7]{$1|\frac{1-\chi_-}{2}$}  (C)
        edge[loop left, left, OliveGreen] node{$1|\frac{1-\chi_+}{2}$}  (D);
    \end{tikzpicture}
}
\caption{Transition diagram for Perturbed Coin Process' q-machine in the discrete Wigner representation. Here, we have defined $\chi_{\pm} = p\pm \sqrt{p(1-p)}$ and use colored edges to easily differentiate the transitions it originates from.}
\label{fig: perturbed coin Wooters QPR transition}
\end{figure}
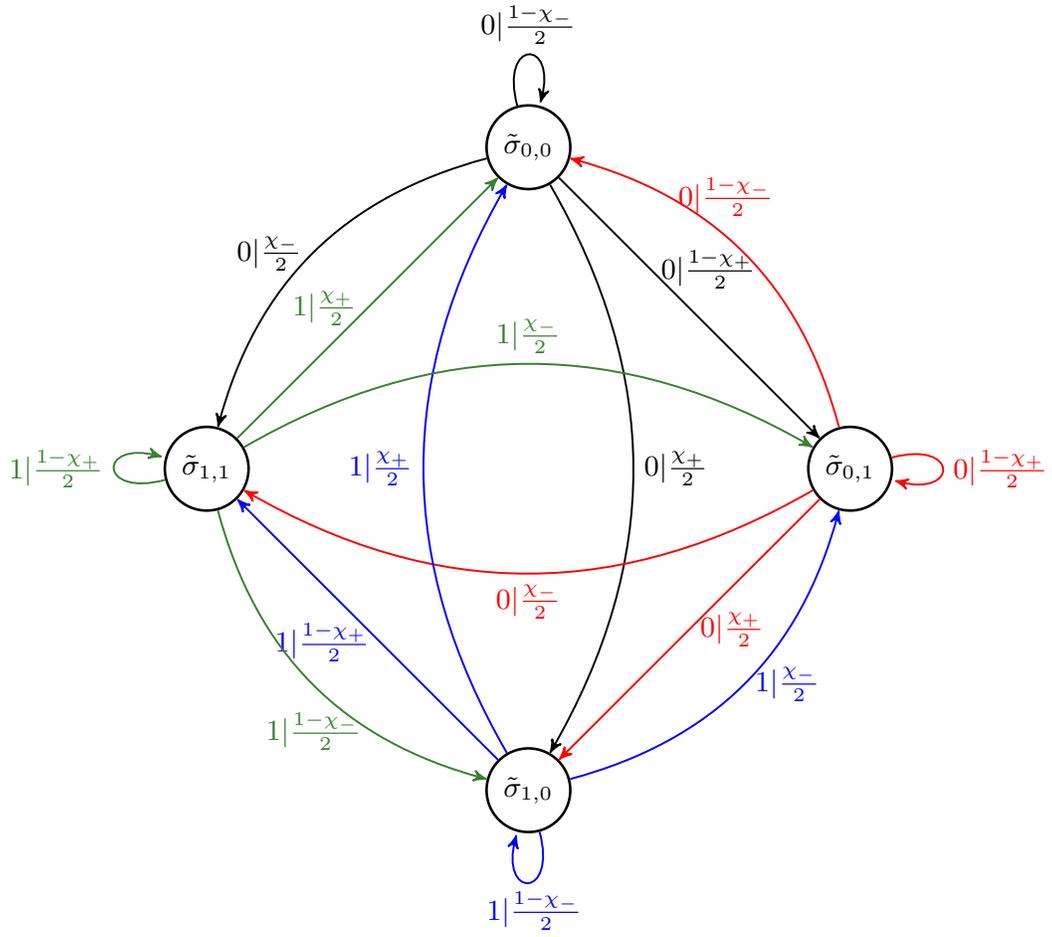

Let us now consider the Perturbed Coin Process again and its q-machine representation. Recall that its quantum state and transition channel on the state are given by
\begin{eqnarray}
\rho &=& \frac{1}{2}\begin{bmatrix}
1 & 2\sqrt{p(1-p)} \\
2\sqrt{p(1-p)} & 1 \end{bmatrix}\, , \\
\kappa[\bullet] &=& \underbrace{K_0 \bullet K_0^\dagger}_{\kappa_0[\bullet]} + \underbrace{K_1 \bullet K_1^\dagger}_{\kappa_1[\bullet]}\, ,
\end{eqnarray}
where $K_0 = \ket{\sigma_0}\bra{0}$, $K_1 = \ket{\sigma_1}\bra{1}$ and $\ket{\sigma_0},\ket{\sigma_1}$ have been defined in the main text. We consider the {\it discrete Wigner representation} \cite{wootters1987wigner} for qubit(s), where $F_\lambda = \frac{1}{2}A_\lambda$, $G_\lambda = A_\lambda$, $\lambda\in \Lambda = \mathbb{Z}_2\times \mathbb{Z}_2$, and the phase point operator $A_\lambda$ reads
\begin{equation}
    A_\lambda \equiv A_{\lambda_1,\lambda_2} = \frac{1}{2}\Big[1 + (-1)^{\lambda_1}Z + (-1)^{\lambda_2}X + (-1)^{\lambda_1+\lambda_2}Y\Big]
\end{equation}
with operators $X,Y,Z$ being Pauli matrices defined in the usual manner. Under this representation, the q-machine above has state and transition matrices given by
\begin{eqnarray}
\rho \longrightarrow \mu^\rho &=& \begin{bmatrix}
\frac{1+2\sqrt{p(1-p)}}{4} & \frac{1-2\sqrt{p(1-p)}}{4} & \frac{1+2\sqrt{p(1-p)}}{4} & \frac{1-2\sqrt{p(1-p)}}{4}
\end{bmatrix}\, ,\label{eq: perturbed coin q-machine state wigner}\\
\kappa_0 \longrightarrow \tau^{\kappa_0} &=& \begin{bmatrix}
\frac{1-p+\sqrt{p(1-p)}}{2} & \frac{1-p-\sqrt{p(1-p)}}{2}  & \frac{p+\sqrt{p(1-p)}}{2} & \frac{p-\sqrt{p(1-p)}}{2} \\
\frac{1-p+\sqrt{p(1-p)}}{2} & \frac{1-p-\sqrt{p(1-p)}}{2} & \frac{p+\sqrt{p(1-p)}}{2} & \frac{p-\sqrt{p(1-p)}}{2} \\
0 & 0 & 0 & 0 \\
0 & 0 & 0 & 0
\end{bmatrix}\, , \label{eq: perturbed coin q-machine output 0 wigner}\\
\kappa_1 \longrightarrow \tau^{\kappa_1} &=& \begin{bmatrix}
0 & 0 & 0 & 0 \\
0 & 0 & 0 & 0 \\
\frac{p+\sqrt{p(1-p)}}{2} & \frac{p-\sqrt{p(1-p)}}{2} & \frac{1-p+\sqrt{p(1-p)}}{2} & \frac{1-p-\sqrt{p(1-p)}}{2} \\
\frac{p+\sqrt{p(1-p)}}{2} & \frac{p-\sqrt{p(1-p)}}{2} & \frac{1-p+\sqrt{p(1-p)}}{2} & \frac{1-p-\sqrt{p(1-p)}}{2}
\end{bmatrix}\, .\label{eq: perturbed coin q-machine output 1 wigner}
\end{eqnarray}
The stationary distribution's entries in $\mu^\rho$ correspond to the internal states $\tilde{\sigma}_{0,0}, \tilde{\sigma}_{0,1} , \tilde{\sigma}_{1,0} , \tilde{\sigma}_{1,1}$, respectively. As can be seen, the state $\rho$ is now represented as a probability distribution $\mu^\rho$ and the elements of the Kraus operators $\kappa_0,\kappa_1$ are now represented as quasiprobabilistic transition matrices $\tau^{\kappa_0},\tau^{\kappa_1}$. Following this, the state transition matrix $\tau^{\kappa} = \tau^{\kappa_0}+ \tau^{\kappa_1}$ is quasi-stochastic matrix --- negativity is found in the state-to-state transitions. This now enable us to visualize the q-machine in terms of directed graph (see Fig. \ref{fig: perturbed coin Wooters QPR transition}) just like $\epsilon$-machines and n-machines. Note that although the stationary distribution here is represented non-negatively, in general a q-machine's stationary distribution is negatively represented (e.g. q-machine of Markov-1 stochastic process \cite{ruebeck2018prediction}). 

 Looking into Fig. \ref{fig: perturbed coin Wooters QPR transition}, one can notice resemblance to the n-machine and how it is constructed. In particular, the states $\{\tilde{\sigma}_{0,0},\tilde{\sigma}_{0,1}\}$ and $\{\tilde{\sigma}_{1,0},\tilde{\sigma}_{1,1}\}$ seems like they are extensions from state $\sigma_0$ and $\sigma_1$ of the $\epsilon$-machine, respectively. This follows from the `state splitting' in Step 1 of the construction. Then the appearance of negativity is inserted in the state-to-state transitions which is the basic idea of Step 2 of the construction. It can be easily seen, however, that although they do not exactly satisfy Eq. \eqref{eq: n-machine construct 1}, they satisfy other properties such as Eqs. \eqref{eq: n-machine construct 3}-\eqref{eq: n-machine construct 2}. This simple example highlights the importance of the n-machine construction in helping us understand the source of nonclassical advantage exhibited by quantum models. As discussed in the main text, we leave more rigorous study of HQMM in quasiprobability representation and its relation with n-machine for future work.

\end{appendix}

\end{document}